\documentclass[twocolumn]{aastex63}

\usepackage{booktabs}
\usepackage{graphicx}
\usepackage{enumitem}

\usepackage{amsmath}
\usepackage{comment}
\usepackage{tabularx}
\usepackage{multirow}
\definecolor{yscol}{rgb}{0.8, 0.6, 1}

\usepackage{natbib}
\usepackage{verbatim}
\usepackage{mathptmx}
\usepackage{url}
\usepackage{numprint}
\usepackage{ulem}

\usepackage{afterpage}

\usepackage{blindtext}
\usepackage{graphicx}
\usepackage{everypage}
\usepackage{environ}

\newcommand{\msun}{\,{\rm M}_\odot}

\newcommand{\s}{\,{\rm s}}
\newcommand{\g}{\,{\rm g}}

\definecolor{yscol}{rgb}{0.8, 0.6, 1}

\definecolor{mycol}{rgb}{0.05, 0.4, 0.1}

\definecolor{delcol}{rgb}{0.6, 0.4, 0.2}

\definecolor{yscol}{rgb}{0.8, 0.6, 1}

\received{April XX, 2023}
\revised{April XX, 2023}
\accepted{May XX, 2023}
\shorttitle{{\it AGORA} Comparison. V: Satellite Galaxies In A Cosmological Zoom-in Simulation}
\shortauthors{{\it AGORA} Collaboration et al.}
\graphicspath{{./}{figures/}}

\begin{document}

\title{The {\it AGORA} High-resolution Galaxy Simulations Comparison Project. V: Satellite Galaxy Populations In A Cosmological Zoom-in Simulation of A Milky Way-mass Halo}

\author[0000-0002-9144-1383]{Minyong Jung}
\correspondingauthor{wispedia@snu.ac.kr}
\affiliation{Center for Theoretical Physics, Department of Physics and Astronomy, Seoul National University, Seoul 08826, Korea}

\author[0000-0002-6299-152X]{Santi Roca-F\`{a}brega}
\altaffiliation{Code leaders}
\correspondingauthor{santi.roca\_fabrega@fysik.lu.se}
\affiliation{Lund Observatory, Department of Astronomy and Theoretical Physics, Lund University, Box 43, SE-221 00 Lund, Sweden}
\affiliation{Departamento de F\'{i}sica de la Tierra y Astrof\'{i}sica, Facultad de Ciencias F\'{i}sicas, Plaza Ciencias, 1, 28040 Madrid, Spain}

\author[0000-0003-4464-1160]{Ji-hoon Kim}
\altaffiliation{Code leaders}
\correspondingauthor{mornkr@snu.ac.kr}
\affiliation{Center for Theoretical Physics, Department of Physics and Astronomy, Seoul National University, Seoul 08826, Korea}
\affiliation{Seoul National University Astronomy Research Center, Seoul 08826, Korea}

\author{Anna Genina}
\altaffiliation{Code leaders}
\affil{Max-Planck-Institut f\"{u}r Astrophysik, Karl-Schwarzschild-Str. 1, D-85748, Garching, Germany}

\author{Loic Hausammann}
\altaffiliation{Code leaders}
\affil{ITS High Performance Computing, Eidgen\"{o}ssische Technische Hochschule Z\"{u}rich, 8092 Z\"{u}rich, Switzerland}
\affil{Institute of Physics, Laboratoire d'Astrophysique, \'{E}cole Polytechnique F\'{e}d\'{e}rale de Lausanne, CH-1015 Lausanne, Switzerland}

\author[0000-0002-7820-2281]{Hyeonyong Kim}
\altaffiliation{Code leaders}
\affiliation{Center for Theoretical Physics, Department of Physics and Astronomy, Seoul National University, Seoul 08826, Korea}
\affiliation{Department of Aerospace Engineering, Seoul National University, Seoul 08826, Korea}

\author{Alessandro Lupi}
\altaffiliation{Code leaders}
\affil{DiSAT, Universit\`a degli Studi dell’Insubria, via Valleggio 11, I-22100 Como, Italy}
\affil{Dipartimento di Fisica ``G. Occhialini'', Universit\`a degli Studi di Milano-Bicocca, I-20126 Milano, Italy}

\author[0000-0001-7457-8487]{Kentaro Nagamine}
\altaffiliation{Code leaders}
\affiliation{Department of Earth and Space Science, Graduate School of Science, Osaka University, Toyonaka, Osaka, 560-0043, Japan}
\affiliation{Kavli IPMU (WPI), University of Tokyo, 5-1-5 Kashiwanoha, Kashiwa, Chiba, 277-8583, Japan}
\affiliation{Department of Physics \& Astronomy, University of Nevada Las Vegas, Las Vegas, NV 89154, USA}

\author[0000-0002-3764-2395]{Johnny W. Powell}
\altaffiliation{Code leaders}
\affil{Department of Physics, Reed College, Portland, OR 97202, USA}

\author{Yves Revaz}
\altaffiliation{Code leaders}
\affil{Institute of Physics, Laboratoire d'Astrophysique, \'{E}cole Polytechnique F\'{e}d\'{e}rale de Lausanne, CH-1015 Lausanne, Switzerland}

\author{Ikkoh Shimizu}
\altaffiliation{Code leaders}
\affil{Shikoku Gakuin University, 3-2-1 Bunkyocho, Zentsuji, Kagawa, 765-8505, Japan}

\author{H\'{e}ctor Vel\'{a}zquez}
\altaffiliation{Code leaders}
\affil{Instituto de Astronom\'{i}a, Universidad Nacional Aut\'{o}noma de M\'{e}xico, A.P. 70-264, 04510, Mexico, D.F., Mexico}

\author{Daniel Ceverino}
\affil{Universidad Aut\'{o}noma de Madrid, Ciudad Universitaria de Cantoblanco, E-28049 Madrid, Spain}
\affil{CIAFF, Facultad de Ciencias, Universidad Aut\'{o}noma de Madrid, E-28049 Madrid, Spain}

\author[0000-0001-5091-5098]{Joel R. Primack}
\affil{Department of Physics, University of California at Santa Cruz, Santa Cruz, CA 95064, USA}

\author{Thomas R. Quinn}
\affil{Department of Astronomy, University of Washington, Seattle, WA 98195, USA}

\author[0000-0001-9695-4017]{Clayton Strawn}
\affil{Department of Physics, University of California at Santa Cruz, Santa Cruz, CA 95064, USA}

\author[0000-0002-5969-1251]{Tom Abel}
\affil{Kavli Institute for Particle Astrophysics and Cosmology, Stanford University, Stanford, CA 94305, USA}
\affil{Department of Physics, Stanford University, Stanford, CA 94305, USA}
\affil{SLAC National Accelerator Laboratory, Menlo Park, CA 94025, USA}

\author{Avishai Dekel}
\affil{Center for Astrophysics and Planetary Science, Racah Institute of Physics, The Hebrew University, Jerusalem 91904, Israel}

\author{Bili Dong}
\affil{Department of Physics, Center for Astrophysics and Space Sciences, University of California at San Diego, La Jolla, CA 92093, USA}

\author[0000-0003-4597-6739]{Boon Kiat Oh}
\affiliation{Department of Physics, University of Connecticut, U-3046, Storrs, CT 06269, USA}
\affiliation{Center for Theoretical Physics, Department of Physics and Astronomy, Seoul National University, Seoul 08826, Korea}

\author{Romain Teyssier}
\affil{Department of Astrophysical Sciences, Princeton University, Princeton, NJ 08544 USA}

\author{the {\it AGORA} Collaboration}
\affiliation{\rm \url{http://www.AGORAsimulations.org}}
\affiliation{\rm The authors marked with * as code leaders contributed to the article by leading the effort within each code group to perform and analyze simulations.} 




\begin{abstract}
We analyze and compare the satellite halo populations at $z\sim2$ in the high-resolution cosmological zoom-in simulations of a $10^{12}\,{\rm M}_{\odot}$ target halo ($z=0$ mass) carried out on eight widely-used astrophysical simulation codes ({\sc Art-I}, {\sc Enzo}, {\sc Ramses}, {\sc Changa}, {\sc Gadget-3}, {\sc Gear}, {\sc Arepo-t}, and {\sc Gizmo}) for the {\it AGORA} High-resolution Galaxy Simulations Comparison Project. We use slightly different redshift epochs near $z=2$ for each code (hereafter ``$z\sim2$') at which the eight simulations are in the same stage in the target halo's merger history.  After identifying the matched pairs of halos between the {\it CosmoRun} simulations and the DMO simulations, we discover that each {\it CosmoRun} halo tends to be less massive than its DMO counterpart.  When we consider only the halos containing stellar particles at $z\sim2$, the number of satellite {\it galaxies} is significantly fewer than that of dark matter halos in all participating {\it AGORA} simulations, and is comparable to the number of present-day satellites near the Milky Way or M31.  The so-called ``missing satellite problem' is fully resolved across all participating codes simply by implementing the common baryonic physics adopted in {\it AGORA} and the stellar feedback prescription commonly used in each code, with sufficient numerical resolution ($\lesssim100$ proper pc at $z=2$).  We also compare other properties such as the stellar mass$-$halo mass relation and the mass$-$metallicity relation.  Our work highlights the value of comparison studies such as {\it AGORA}, where outstanding problems in galaxy formation theory are studied simultaneously on multiple numerical platforms.
\end{abstract}





\section{Introduction}

Studied extensively by cosmologists, the $\Lambda$-Cold Dark Matter ($\Lambda$CDM)  model is considered the standard model of Big Bang cosmology, encompassing dark energy and dark matter.
However, there is a certain tension between theory and observed galaxies, especially on a small scale \citep[for reviews, see e.g.,][]{2017ARA&A..55..343B, 2017Galax...5...17D}. For example, the observed number of dwarf galaxies around the Local Group is significantly fewer than that of the dark matter halos found in $N$-body simulations when compared based on their circular velocity. This so-called ``missing satellite problem'' is one of the long-standing challenges of the contemporary $\Lambda$CDM model \citep{1993MNRAS.264..201K, 1999ApJ...522...82K, 1999ApJ...524L..19M, 2002MNRAS.333..156B}.
Reproducing satellite galaxies and small-scale substructures in a simulation within the $\Lambda$CDM framework is a nontrivial task because it requires high numerical resolution and sophisticated baryonic physics.

The mismatch between the theory and the observation on a small scale has motivated a great deal of theoretical modeling such as the warm dark matter \citep[WDM; e.g.,][]{2001ApJ...556...93B}, fuzzy dark matter \citep[e.g.,][]{2000PhRvL..85.1158H}, and self-interacting dark matter \citep[SIDM; e.g.,][]{2000PhRvL..84.3760S}. 
By suppressing the small-scale matter power spectrum in the early universe and/or stimulating halo disruptions at later times, these alternative dark matter models have shown to reduce the number of subhalos around the Milky Way (MW)-mass halos \citep[e.g.,][]{2011arXiv1109.6291D, 2021ApJ...920L..11N}.

On the other hand, it is possible that baryonic processes could suppress the formation of some dwarf galaxies or make them difficult to observe, which could explain the missing satellite problem \citep{2010ApJ...709.1138D, 2013ApJ...765...22B,2014ApJ...786...87B,2016MNRAS.457.1931S, 2016ApJ...827L..23W, 2021ApJ...906...96A}. 
In such cases, dark matter halos may still exist, but may not have formed visible dwarf galaxies due to the effects of baryonic physics. 
This is a possible solution to the missing satellite problem within the framework of the $\Lambda$CDM paradigm. 
In addition, many authors have shown that low-mass halos could be easily disrupted by baryon-induced physics such as cosmic reionization, tidal stripping, ram pressure stripping, and stellar feedback \citep{2016MNRAS.458.1559Z, 2018MNRAS.478..548S}.
In the {\it Latte} simulations with the FIRE star formation and feedback model, the dwarf galaxy population near the MW/M31-mass halo was found to agree well with the observed population in the Local Group \citep{2016ApJ...827L..23W}. 
Meanwhile, the ``Mint'' resolution DC Justice League suite of MW-like zoom-in simulations showed that the number of satellite galaxies matches the observed population of the dwarf galaxies around MW-sized galaxies down to the ultrafaint dwarf regime \citep[UFD;][]{2021ApJ...906...96A}. 
Some studies have also shown that $\Lambda$CDM simulations can reproduce the radial distribution of MW satellites \citep{2018MNRAS.473.4392S, 2019MNRAS.487.1380G, 2020MNRAS.491.1471S}. 
Moreover, the number of observed faint galaxies has increased recently \citep[for reviews, see][]{2019ARA&A..57..375S}, which partially mitigates the missing satellite problem. 
These findings suggest that the missing satellite problem is very close to being solved.
In fact, some researchers such as, \cite{2018PhRvL.121u1302K} and \cite{2022NatAs.tmp..130S}, argue that the problem is resolved.

Ideally, we would then expect the satellite galaxy populations to be consistent regardless of the simulation code utilized. 
Nevertheless, due to differences in the inherent properties of the simulations such as the adopted physics models and the implementations of the gravity solver, discrepancies may arise between codes \citep{2005ApJS..160....1O, 2008CS&D....1a5003H}. 
For example, \cite{2016MNRAS.458.1096E} studied subhalos and galaxies in a galaxy cluster produced by multiple simulation codes, and found that in dark matter-only (DMO) simulations, the population and properties of subhalos show good agreements across code platforms. 
Nevertheless, they also discovered that the codes produce significantly different galaxy populations when baryonic physics models were included. 
While they found both similarities and disparities in the galaxy population, the comparison of dwarf galaxy populations ($M_{\rm halo} < 10^{10} \msun$) was not feasible due to the limited resolution of their simulations. 
Indeed, there is an urgent need for controlled comparisons of the dwarf galaxy populations produced by different simulation codes. 
Such comparisons will be essential to understand the robustness of the satellite galaxy populations predicted in the simulations, and how sensitive they are with respect to the specific numerical methods and assumptions adopted in the simulations.  

The {\it AGORA} High-resolution Galaxy Simulations Comparison Project ({\it Assembling Galaxies of Resolved Anatomy}) has aimed at collectively raising the predictive power of numerical galaxy formation simulations, by comparing high-resolution galaxy-scale calculations across multiple code platforms, using a DMO galaxy formation simulation \citep[][hereafter Paper I]{2014ApJS..210...14K_short}, an idealized disk galaxy formation simulation \citep[][hereafter Paper II]{2016ApJ...833..202K_short}, and a fully cosmological zoom-in galaxy formation simulation \citep[][hereafter Papers III and IV]{2021ApJ...917...64R, 2022Santi}. 
In this paper, we analyze the satellite halos around the target MW-like halo in the {\it AGORA ``CosmoRun''} simulation suite introduced and studied in Papers III and IV. 
Specifically, we compare the eight hydrodynamic {\it CosmoRuns} and eight DMO simulations, all performed with the state-of-the-art galaxy simulation codes widely used in the numerical galaxy formation community, and study the populations of their satellite halos and galaxies.  
We choose slightly different redshift epochs near $z=2$ for each code in order to compare the runs at the same dynamical stage in the target halo's evolution history (see Section \ref{cosmorun} for details). 
We then compare the number of satellite halos in {\it CosmoRuns} with its counterpart in the DMO simulations. 
We also explore the consistency between the codes in other properties of satellite galaxies, including the stellar mass$-$halo mass relation as well as the mass$-$metallicity relation.

This paper is organized as follows. 
Section \ref{sec:methodology} describes the {\it AGORA} {\it CosmoRun} and the DMO simulation, as well as the definition of a satellite halo. 
In Section \ref{sec:results}, the satellite halo and galaxy populations in the {\it CosmoRuns} are presented in comparison with those in the DMO runs. 
In Section \ref{sec:discussion}, based on our results we predict the satellite galaxy population at $z\sim0$, and test inter-code convergence in other satellite properties.
Finally, we conclude the paper in Section \ref{sec:conclusion}.


\section{Methodology}
\label{sec:methodology}

\subsection{The {\it AGORA ``CosmoRun''} Simulation Suite}
\label{cosmorun}

The {\it CosmoRun} described in Paper III is a suite of high-resolution cosmological zoom-in simulations of a MW-mass halo ($10^{12}\,{\rm M}_{\odot}$ at $z=0$) on multiple code platforms.\footnote{For publicly available datasets, visit \url{http://www.AGORAsimulations.org} or \url{http://flathub.flatironinstitute.org/agora}.}   
The simulations analyzed herein started from a cosmological initial condition at $z = 100$ and reached $z \lesssim 2$. 
The adopted cosmological parameters are $\Omega_{\Lambda}$ = 0.728, $\Omega_{\rm matter} = 0.272$, $\Omega_{\rm DM}=0.227$, $\sigma_{8} = 0.807$, $n_{\rm s} = 0.961$, and $h = 0.702$.
The code groups participating in this particular comparison encompass both particle-based and mesh-based codes:  {\sc Art-I}, {\sc Enzo} and {\sc Ramses} are mesh-based codes whereas {\sc Changa}, {\sc Gadget-3}, {\sc Gear}, {\sc Arepo-t} and {\sc Gizmo} are particle-based codes.\footnote{We classify the SPH codes ({\sc Changa}, {\sc Gadget-3}, {\sc Gear}) and the arbitrary Lagrangian-Eulerian codes ({\sc Arepo} and {\sc Gizmo}) as particle-based codes.} Galaxy formation has been studied using both approaches, each with its own advantages and disadvantages. 
After a series of calibration steps, all the codes in Paper III reached an overall agreement in the stellar properties of the target halo, and in its mass assembly history. 
The final {\it CosmoRun} suite includes common baryonic physics modules in {\it AGORA} such as the {\sc Grackle} radiative gas cooling \citep{2017MNRAS.466.2217S}, cosmic ultraviolet background radiation \citep{HaardtMadau12}, and star formation, as well as the code-dependent physics including --- most notably --- stellar feedback prescriptions. 
Both code-independent and code-dependent physics implemented in each code are explained in great detail in Paper III (some in Paper II). We update the two models from Paper III, {\sc Art-I} and {\sc Changa}, to include weaker stellar feedback. In {\sc Art-I}, we change the condition for the minimum time step at high redshifts to achieve better convergence in the halo growth history. We also incorporate a new model using the {\sc Arepo} code into our analysis. We refer to this as {\sc Arepo-t}, which represents the {\sc Arepo} code with thermal feedback. The differences between the old and new {\sc Art-I} and {\sc Changa} models, as well as the details of the {\sc Arepo-t} model, are illustrated in Paper IV.\footnote{In the analysis presented in Section \ref{galaxy_z0}, we used the older {\sc Art-I} model, labeled as {\sc Art-I} (old), which is described in Paper III, because the new model has not reached $z\lesssim 1$. The results with both models are mostly consistent.}

The gravitational force softening length for the particle-based codes in the highest-resolution region is 800 comoving pc until $z=9$ and 80 proper pc afterward.
Meanwhile, the finest cell size of the mesh-based codes is set to 163 comoving pc, or 12 additional refinement levels for a $128^3$ root resolution in a $(60 \,\,\,{\rm comoving} \,\,h^{-1}\,{\rm Mpc})^3$ box.   
A cell is adaptively refined into 8 child cells on particle over-densities of 4. 
For details on runtime parameters, we refer the readers to Paper III.

While all the {\it AGORA} {\it CosmoRun} simulations were calibrated to produce similar stellar masses in the host halo by $z=4$ (see Section 5.4 and Figure 12 in Paper III), we find that the host halo in some codes' {\it CosmoRun} are at a different stage in its dark matter accretion history from others' at $z=2$.
This is likely due to the inter-code ``timing discrepancy'' (see Section 5.3 in Paper I for more information). 
Because the halos in different codes are at different evolutionary stages, the satellite halo abundances are also different among the {\it CosmoRun}. 
To resolve this timing discrepancy, we have created a merger tree for each code and selected an epoch near $z=2$ (hereafter called ``$z\sim 2$'') for each code so that the target halo is in the same stage in its merger history (for more information, see Paper IV). 
The list of epochs for each code used for the present paper is in Table \ref{tab:redshifts}. 
Snapshots of the {\it CosmoRun} simulations at $z\sim2$ are shown in Figure \ref{fig:halo_plot}.
\begin{table}[]
\centering
\vspace{2mm}
\begin{tabular}{ccc}
\hline
\multirow{2}{*}{Code} & \multicolumn{2}{c}{Redshift epoch}\\\cline{2-3}
& {\it CosmoRun} & Dark matter-only (DMO) run\\\hline\hline
{\sc Art-I}  & 1.85    & 2.18    \\
{\sc Enzo}   & 2.29    & 2.15   \\
{\sc Ramses}  & 2.21  & 2.12    \\
{\sc Changa}   & 2.08   & 2.09   \\
{\sc Gadget-2/3}   & 2.13 & 2.05   \\
{\sc Gear}   & 1.88  & 1.87     \\
{\sc Arepo-t}  & 1.98 & 2.11     \\
{\sc Gizmo}  & 2.02 & 2.11     \\\hline
\end{tabular}
\vspace{3mm}
\caption{The redshift epoch selected for each code to be analyzed in this paper.  At these epochs, the eight {\it CosmoRuns} are in the same stage in the target halos' merger history.  See Section \ref{cosmorun} for details.}
\vspace{-2mm}
\label{tab:redshifts}
\end{table}

\subsection{The Dark Matter-Only (DMO) Simulations}
\label{dmorun}

\begin{figure*}
    \vspace{0mm}
    \centering
    \includegraphics[width = 1\linewidth]{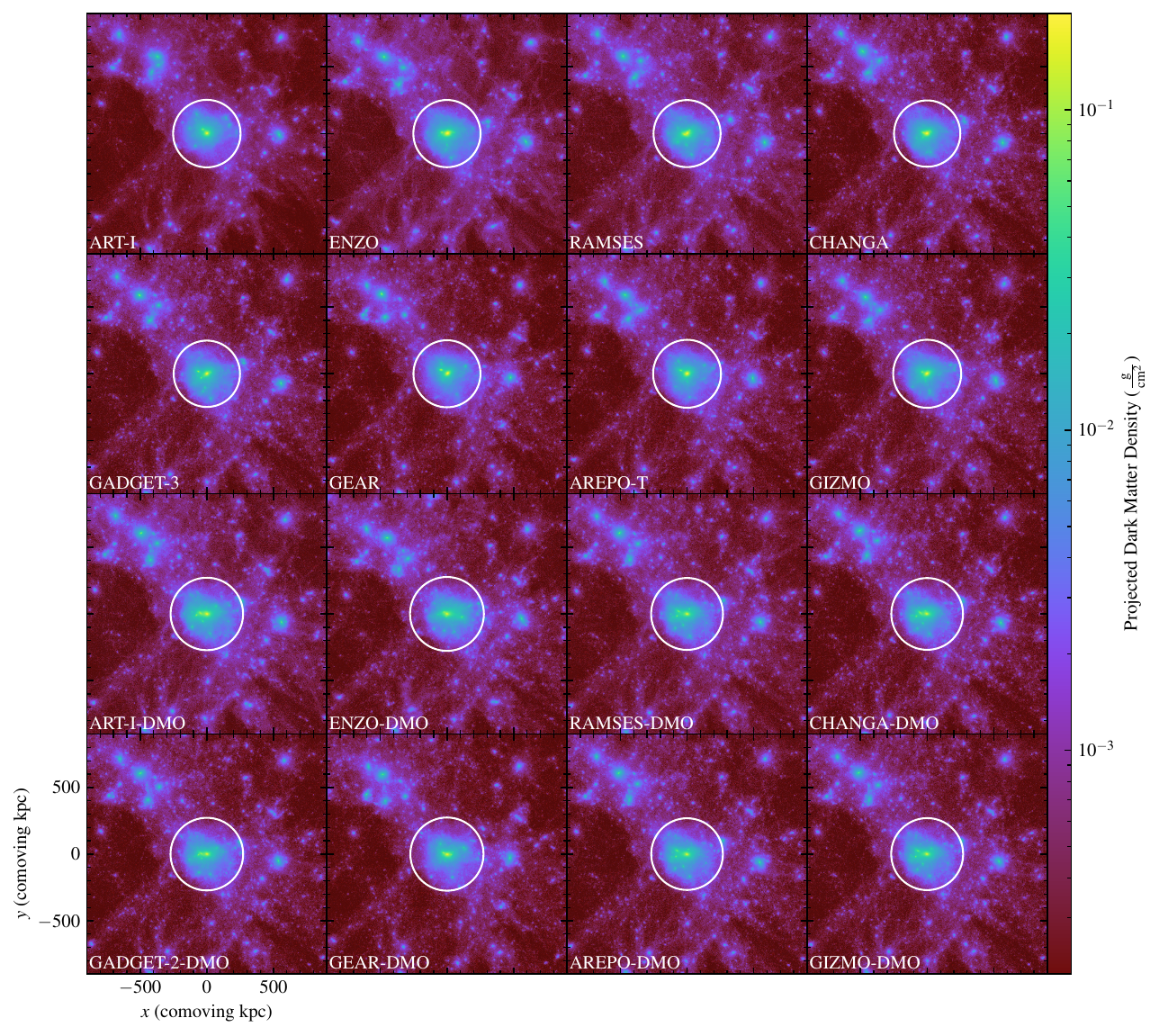}
    \vspace{-5mm}
    \caption{The dark matter surface densities at $z \sim 2$ (the exact redshift in each code in Table \ref{tab:redshifts}) for eight hydrodynamic {\it ``CosmoRun''} simulations and eight dark matter-only (DMO) simulations, projected through a 1.8 comoving Mpc thick slab, with the target host halo's virial radius $R_{\rm vir}$ drawn in a {\it white circle}. See Section \ref{sec:methodology} for more information on these simulations.         Simulations performed by:  Santi Roca-F\`{a}brega ({\sc Art-I}, {\sc Ramses}, and {\sc Art-I-dmo}), Ji-hoon Kim ({\sc Enzo}), Johnny Powell and H\'ector Vel\'azquez ({\sc Changa} and {\sc Changa-dmo}), Kentaro Nagamine and Ikkoh Shimizu ({\sc Gadget-3}), Loic Hausammann and Yves Revaz ({\sc Gear} and {\sc Gear-dmo}), Anna Genina ({\sc Arepo-t}, and {\sc Arepo-dmo}), Alessandro Lupi and Bili Dong ({\sc Gizmo}), Hyeonyong Kim ({\sc Enzo-dmo}, {\sc Ramses-dmo}, {\sc Gadget-2-dmo}, and {\sc Gizmo-dmo}). Note that the mean dark matter surface densities in DMO runs are $\Omega_{\rm matter}/\Omega_{\rm DM} = 1.20$ times higher since it includes the contribution from baryons. The high-resolution versions of this figure and article are available at the Project website, \url{http://www.AGORAsimulations.org/}.}

    \label{fig:halo_plot}
\end{figure*}

In order to investigate the role of baryonic physics adopted for {\it AGORA} in the satellite halo population, we have also performed DMO simulations using the same zoom-in initial condition generated with {\sc Music} \citep{2011MNRAS.415.2101H} but with no gas component. Accordingly, the mass of the dark matter particles in the DMO runs is $\Omega_{\rm matter}/\Omega_{\rm DM} = 1.20$ times heavier than that in the {\it CosmoRun}.
While Paper I found that the dark matter properties and the satellite halo populations are nearly identical across all participating codes in {\it AGORA}, there remained a systematic discrepancy in the satellite halo populations in the low-mass end. 
Therefore, we have employed all eight codes in {\it AGORA} to run DMO simulations to check their consistency.\footnote{In terms of the gravity solver for collisionless components, {\sc Gadget-2} (latest version in 2011) and {\sc Gadget-3} \citep[first introduced in][]{2008MNRAS.391.1685S} are identical for our purpose, and will produce practically identical results in the DMO runs.}
Snapshots of these DMO runs $z \sim 2$ are also included in Figure \ref{fig:halo_plot}.
The runtime parameters governing the collisionless dynamics in the DMO runs are set to be identical to those used for the {\it CosmoRun}. 

\begin{figure*}
    \vspace{0mm}
    \centering
    \includegraphics[width = 1\linewidth]{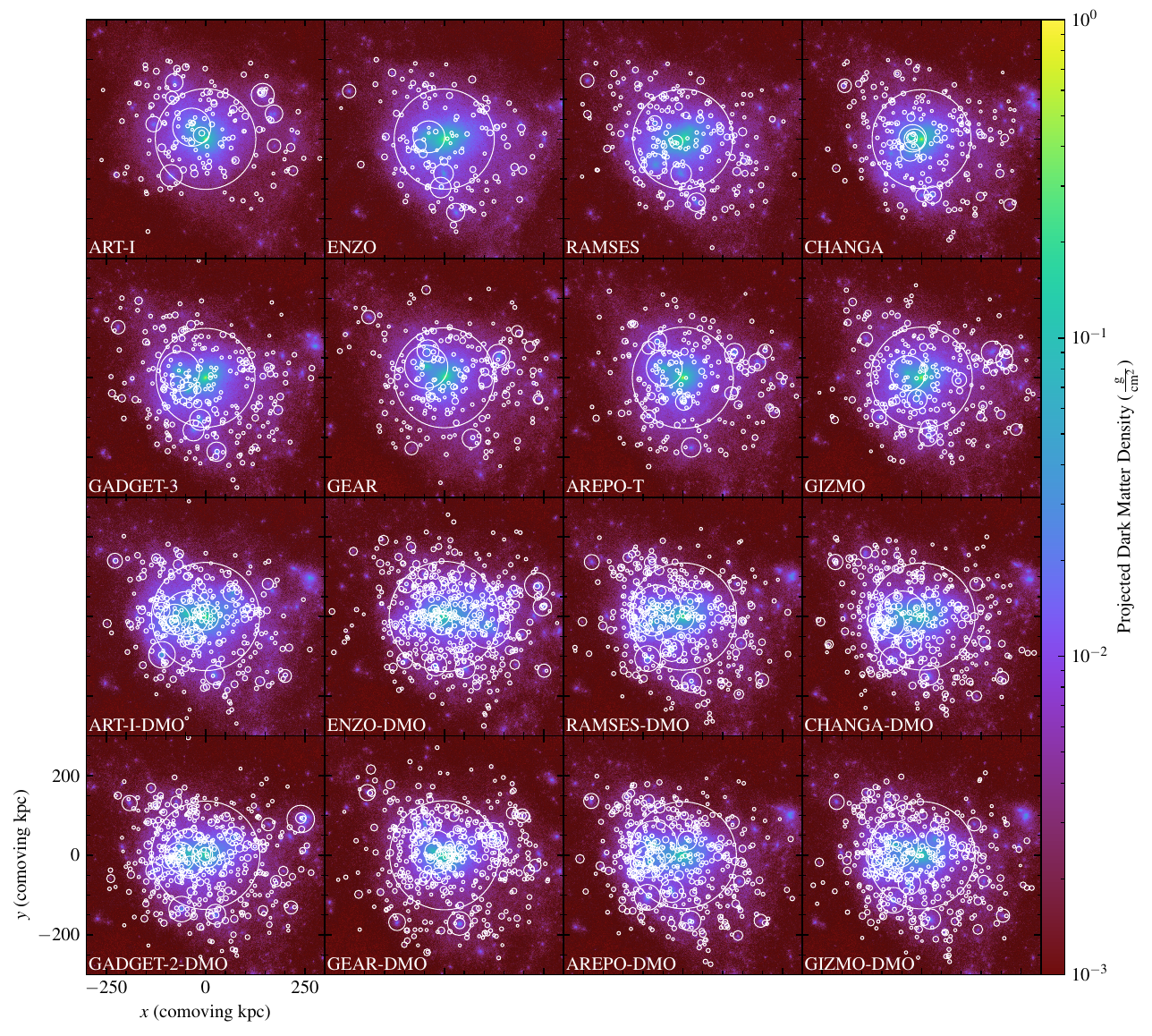}
    \vspace{-5mm}
    \caption{The dark matter surface densities at $z\sim2$ with the halos identified by the {\sc Rockstar} halo finder drawn in {\it white circles} whose radii indicate $0.5R_{\rm vir}$. Only the halos located within 300 comoving kpc from the host halo's center and those more massive than $10^7h^{-1}\msun$ in dark matter are drawn. Readers can readily see that the DMO runs have more satellite halos than the {\it CosmoRuns}. See Section \ref{abundance} for more information.}

    \label{fig:halo_plot_600}
\end{figure*}

\subsection{Halo Finding}
\label{halo_finding}

Halos in the {\it CosmoRun} and the DMO runs are identified with the {\sc Rockstar} halo finder \citep{2013ApJ...762..109B} using only the highest-resolution dark matter particles (i.e., not stellar or gas particles). 
We further narrow down the identified halos to {\it satellite halos} using the following criteria: {\it (i)} it must reside within 300 comoving kpc from the target host halo (100 proper kpc at $z=2$; similar to the virial radius, $R_{\rm vir}$, of our host halo at $z=0$), and {\it (ii)} it must be more massive than $10^7 h^{-1} \msun$ in dark matter (equivalent to 45 dark matter particles in the DMO runs).\footnote{Note that there is no velocity criteria or requirement when identifying satellites. Therefore, some halos may be counted as satellites despite not being gravitationally bound to the host halo.} 
We follow \cite{1998ApJ...495...80B} definition of virial radius and mass.

For our analysis in Sections \ref{stellar_halos} and \ref{sec:discussion}, we assign a stellar particle to a halo following the process in \cite{2020MNRAS.491.1471S}. 
We first identify all stellar particles located within $0.8R_{\rm vir}$ from the halo, with their velocities relative to the halo less than twice the halo's maximum circular velocity.
We then calculate the radius that encompasses 90\% of the stellar particles ($R_{90}$) and the stellar velocity dispersion ($\sigma_{\rm vel}$). 
To further refine our selection, we narrow down the stellar particle list to those satisfying two more conditions: (1) they are located within $1.5\,R_{90}$ from the center of mass of the halo {\it and} stellar particles, and (2) their velocities relative to the halo is less than $2\sigma_{\rm vel}$.
We then iterate the analysis, recalculating $R_{90}$ and $\sigma_{\rm vel}$ for the selected member particles until they converge within 99\% of the previous values. We start from the most massive satellite halos to lower ones, making sure not to reassign stellar particles that have already been allocated.
Finally we define satellite ``galaxies'' as those whose stellar masses are at least six times the approximate mass resolution of stellar particles (i.e., $M_{\rm star} > 6m_{\rm gas,\,IC} = 2.38\times 10^5 h^{-1}\msun$; see Section 3.1 of Paper III).


\vspace{1mm}
\section{Results}
\label{sec:results}

\subsection{Satellite Halo Populations At $z\sim 2$}
\label{abundance}

Figure \ref{fig:halo_plot_600} shows the dark matter surface density plots in a (600 comoving kpc)$^2$ box, with the target host halo and the satellite halos drawn in white circles (whose radii indicate half the virial radii, $0.5 R_{\rm vir}$). 
One can already observe that the eight hydrodynamic {\it CosmoRuns}  have produced similar numbers of dark matter halos with similar $R_{\rm vir}$'s for the host halo.    
Readers can also see that the DMO runs clearly have more satellite halos than the {\it CosmoRuns}.

\begin{figure*}
    \vspace{1mm}
    \centering
    \includegraphics[width=0.91\linewidth]{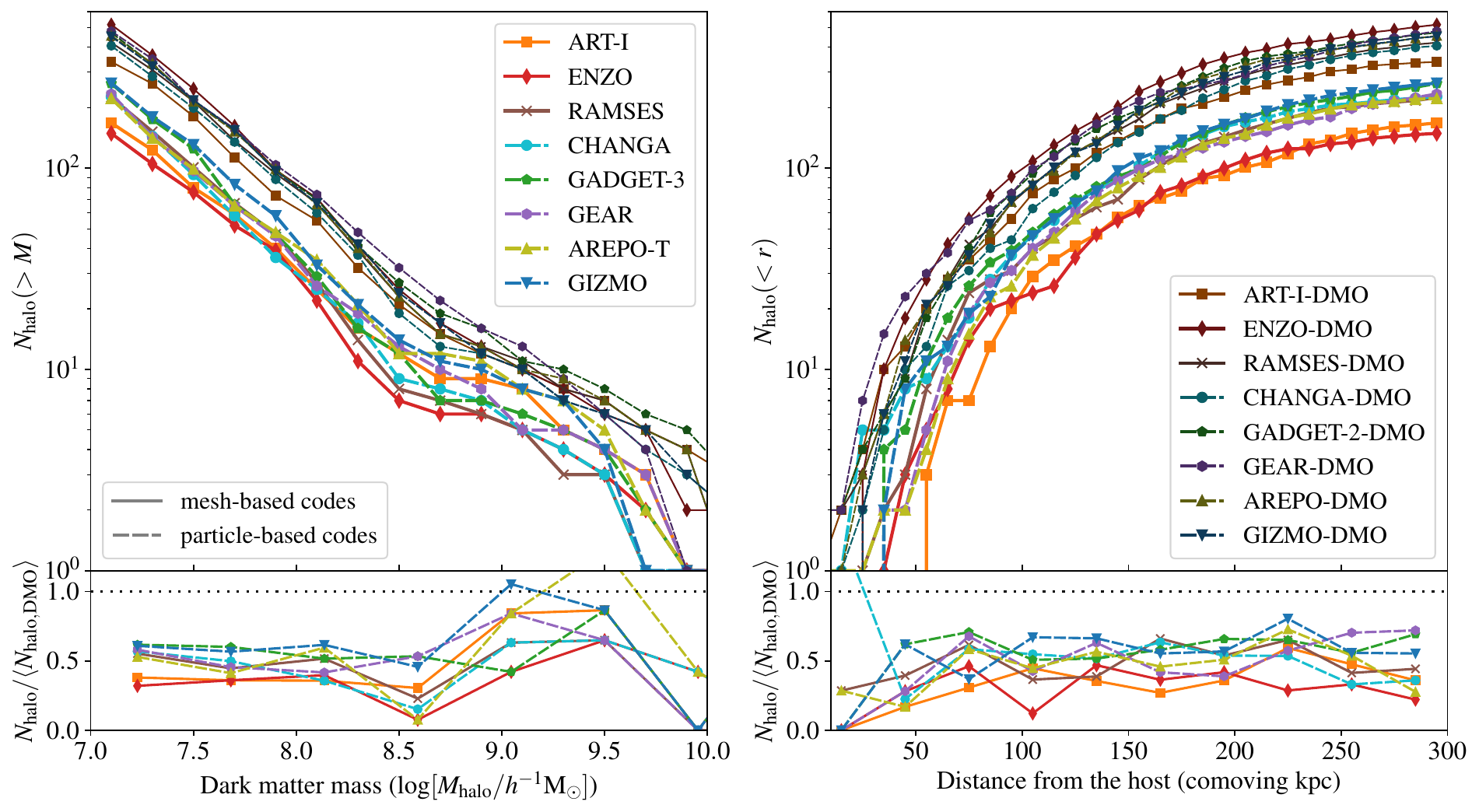}
    \vspace{-1mm}    
    \caption{The cumulative number of satellite halos at $z \sim 2$ in their dark matter mass, $N_{{\rm halo}} (>M)$ ({\it left}), and in radial distance from the host halo's center, $N_{{\rm halo}} (<r)$ ({\it right}).  
    We distinguish mesh-based codes ({\it solid lines}) and particle-based codes ({\it dashed lines}) with different line styles, and {\it CosmoRuns} and DMO runs with different brightness. 
    The {\it bottom panels} display the ratio of the number of the CosmoRun’s satellite halos to that in the DMO run (the mean value of the eight DMO runs) in each mass/radius bin.
    All hydrodynamic {\it CosmoRuns} have fewer satellite halos than the DMO runs do across all halo masses and radii.
    See Section \ref{abundance} for more information.
    }
    \label{fig:mass_function_z2}
    \vspace{2mm}
\end{figure*}

To quantitatively study the differences in the participating simulations, in Figure \ref{fig:mass_function_z2} we plot the cumulative number of satellite halos at $z \sim 2$ in their dark matter mass, $N_{{\rm halo}} (>M)$ (left panels), and in radial distance from the host halo's center, $N_{{\rm halo}} (<r)$ (right panels).
It is worth noting several points: 
\begin{itemize}

\item First, we find that all eight hydrodynamic {\it CosmoRuns} have fewer satellite halos than the DMO runs do across all halo masses and radii.
In the halo mass function (left panels of Figure \ref{fig:mass_function_z2}), the numbers of satellite halos in all {\it CosmoRuns} are systematically fewer than those in the DMO runs by a factor of $\sim\,2$ for $M_{\rm halo}$ (halo dark matter mass) $ < 10^{8.5} h^{-1} \msun$.
To put it differently, the ratios of the number of the {\it CosmoRun} satellite halos to that in the DMO run (the mean number of halos in the eight DMO runs) in each mass bin, $N_{\rm halo}/\langle N_{\rm halo,\,DMO} \rangle$, is $\sim\,$0.5 (bottom left panel).

\item Second, the ratio of the satellite halos' {\it radial} distribution function in the {\it CosmoRun} to that in the DMO run, $N_{\rm halo}/\langle N_{\rm halo,\,DMO} \rangle$, tends to become small --- often zero --- in the bin closest to the host halo's center, $r < 40 \,\,\,{\rm comoving \,\,kpc}$ (bottom right panel of Figure \ref{fig:mass_function_z2}). 
This implies that the causes of the deficit --- the effect of baryonic physics which we will explore in depth in Section \ref{over_time} --- have a stronger influence near the host halo's center. This is consistent with the findings of earlier studies \citep[e.g.,][]{2014ApJ...786...87B, 2016ApJ...827L..23W, 2017MNRAS.467.4383S, 2017MNRAS.471.1709G, 2019MNRAS.487.4409K}.

\item The satellite halo populations in the eight DMO runs are slightly different but are in general agreement with one another in both mass and space (upper panels of Figure \ref{fig:mass_function_z2}; lines with a reduced stroke width and darker colors). 
Among the DMO runs, no systematic difference exists between the mesh-based and particle-based codes, a result somewhat different from the earlier studies \cite[e.g.,][]{2005ApJS..160....1O, 2008CS&D....1a5003H} or from our findings in Paper I.\footnote{In Paper I, systematic difference between particle-based and mesh-based codes at the low-mass end was observed.  It was because mesh-based codes tend to have coarser force resolution as they attempt to resolve minute density fluctuations in the outskirts of the target halo at high redshift, resulting in a difference in the abundance of low-mass satellite halos.  Note also that the analysis in Paper I was carried out with the {\sc HOP} halo finder, a different choice from  the {\sc Rockstar} halo finder for what is presented here, which could produce different numbers of halos identified.}
Instead, {\sc Enzo-dmo} has a slightly higher number of halos compared to the particle-based codes, and {\sc Ramses-dmo} is in the middle of the pack of particle-based codes.
The numerical resolution in the highest-resolution region of the {\it CosmoRun} --- and correspondingly, our new DMO runs --- is chosen to resolve the interstellar medium (ISM) and the star-forming regions in them (i.e., $\lesssim 100$ proper pc at all times between $z=100$ and 2).
The high resolution in our simulation suite might have been sufficient for {\sc Enzo-dmo} and {\sc Ramses-dmo} to alleviate any discrepancy previously observed in the satellite halo population between the particle-based and mesh-based DMO runs. 
Note that {\sc Art-I-dmo} seems to have had a slightly harder time fully resolving the outskirts of our target halo.\footnote{Note that runtime parameters are not chosen to match the refinement structure between the  {\it CosmoRun} and the DMO run.  In a typical mesh-based DMO run, cells are adaptively refined only by dark matter mass, whereas in a {\it CosmoRun} they are refined not only by dark matter mass but also by baryon mass and others.  As a result, {\sc Art-I} and {\sc Art-I-dmo} may have different refinement structure, especially in the outskirts of the target halo.}
Even so, the difference is not as severe as what was seen in the previous studies.  

\item Two {\it CosmoRuns}, {\sc Art-I} and {\sc Enzo}, have smaller satellite halo populations than the rest of the participating codes do, especially in the low-mass end ($M_{\rm halo} < 10^{7.5} h^{-1} \msun$) and in the outskirts of the host halo ($r > 200\,\,\, {\rm comoving \,\,kpc}$).
And the number of halos in the three mesh-based {\it CosmoRuns} tends to be lower in the range of halo masses $M_{\rm halo} \sim 10^{8.5} h^{-1}\msun$ and radial distances $\left[100, \,150\right]$ comoving kpc.
The inter-code difference among the {\it CosmoRuns}, which their counterpart DMO runs do not exhibit, should be attributed to how the same (or similar) baryonic physics are treated differently in the two hydrodynamics approaches.

\end{itemize}

One of the most notable findings among the above is that all hydrodynamic {\it CosmoRuns}  have produced fewer satellite halos than the DMO runs have by $z\sim 2$ across all halo masses and radii. 
We further study in Section \ref{stellar_halos} that the so-called ``missing satellite problem'' \citep[over-abundance of satellite halos in simulations;][]{1993MNRAS.264..201K, 1999ApJ...522...82K, 1999ApJ...524L..19M, 2002MNRAS.333..156B} could be easily resolved  in all participating codes simply by implementing the baryonic physics adopted for {\it AGORA} in simulations with sufficient numerical resolution ($\lesssim 100$ proper pc at $z=2$) by examining the satellite {\it galaxy} populations around the target host halo.
For now, in Sections \ref{over_time} and \ref{matching_pairs} we focus on the causes of differences seen in Figure \ref{fig:mass_function_z2} --- between the satellite {\it halos} in the {\it CosmoRuns} and in the DMO runs.   

\begin{figure*}
    \vspace{1mm}
    \centering
    \includegraphics[width=0.91\linewidth]{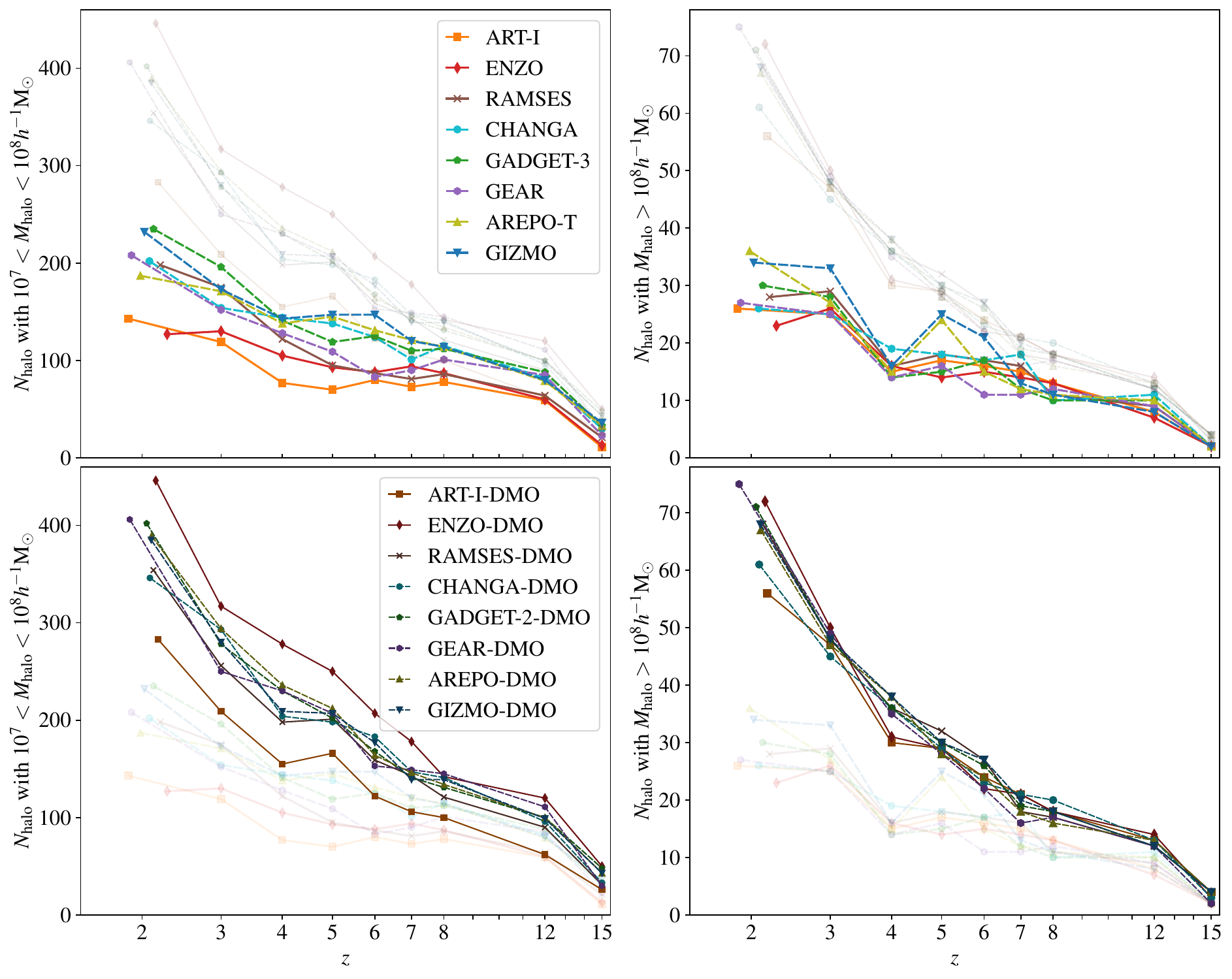}
    \vspace{-1mm}   
    \caption{The evolution of the number of satellite halos across cosmic time, for two different dark matter mass bins, $10^7h^{-1}\msun < M_{\rm halo}<10^8 h^{-1}\msun$ ({\it left}) and $M_{\rm halo}>10^8h^{-1}\msun$ ({\it right}). 
    We only count halos that reside within 300 comoving kpc from the target host halo. 
    The {\it top panels} highlight the {\it CosmoRuns} with the DMO runs shown at reduced opacity, while the {\it bottom panels} emphasize the DMO runs with the {\it CosmoRuns} at reduced opacity.   
    All hydrodynamic {\it CosmoRuns} have fewer satellite halos than the DMO runs do in both mass bins at nearly all redshifts.    
    The mesh-based {\it CosmoRuns} tend to host slightly fewer lower-mass satellite halos than the particle-based {\it CosmoRuns} do at most redshifts ({\it top left panel}).
    See Section \ref{over_time} for more information.     
    }
\label{fig:num_halo_by_z}
    \vspace{2mm}    
\end{figure*}

\vspace{1mm}
\subsection{Evolution of Satellite Halo Populations} 
\label{over_time}

We now study the halo populations at multiple epochs from $z=15$ to $\sim2$ to understand and discriminate various causes that have affected the satellite halo populations.  
Figure \ref{fig:num_halo_by_z} shows the evolution of the number of satellite halos in two different mass bins, $ 10^7 h^{-1} \msun  < M_{\rm halo} < 10^8 h^{-1} \msun$ (left panels) and $M_{\rm halo} > 10^8 h^{-1}  \msun$ (right panels). 
Readers can notice that the hydrodynamic {\it CosmoRun} simulations and the DMO runs show systematic differences in both mass bins at nearly all redshifts. 
In the left panels of Figure \ref{fig:num_halo_by_z}, one may spot the systematic disagreements between the DMO runs, the group of particle-based {\it CosmoRuns}, and the group of mesh-based {\it CosmoRuns}, already at $z=15$.  
From $z=8$ to 4, the numbers of halos in the {\it CosmoRuns} barely grows in both mass bins, while those in the DMO runs increase steadily. 
The number of halos in the {\it CosmoRuns} in the higher-mass bin (right panels; $M_{\rm halo} > 10^8 h^{-1}  \msun$) remain approximately constant from $z=3$ to $\sim 2$, whereas those in the DMO runs continue to increase. 
In the meantime, just as in Figure \ref{fig:mass_function_z2}, there exists disagreement between the particle-based codes (colored dashed lines in the top panels of Figure \ref{fig:num_halo_by_z}) and the mesh-based codes (colored solid lines), especially in the lower-mass bin (top left panel; $ 10^7 h^{-1} \msun  < M_{\rm halo} < 10^8 h^{-1} \msun$).

Now we investigate various causes for these differences in time:

\begin{figure*}
     \vspace{1mm}
    \centering
    \includegraphics[width=0.95\linewidth]{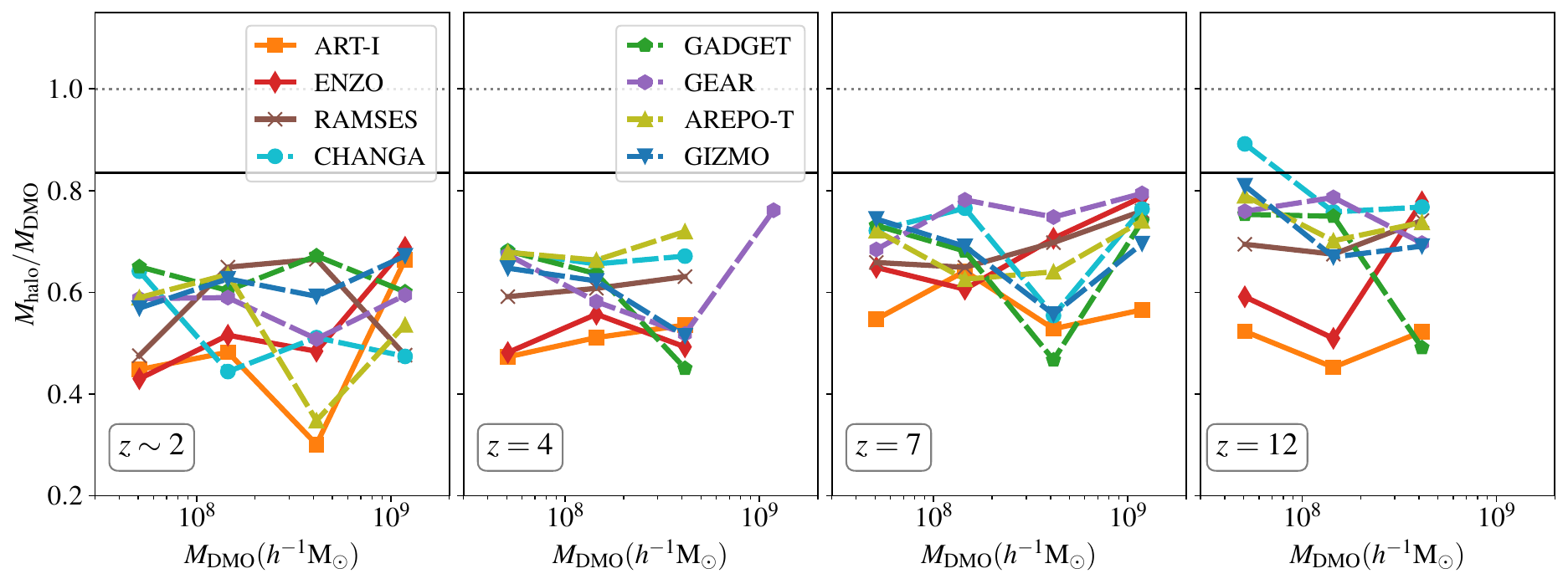}
     \vspace{-1mm}
     \caption{The ratio of the dark matter mass of an individual halo in the {\it CosmoRun} to that of its DMO counterpart ($M_{\rm halo} / M_{\rm DMO}$). 
     Each symbol represents a {\it CosmoRun} -- DMO run pair at four different epochs, from $z = 12$ to $\sim2$. 
     A {\it thick solid line} is the median value in each mass bin for each {\it CosmoRun}. 
     The {\it solid horizontal line} denotes $\Omega_{\rm DM}/\Omega_{\rm matter} = 0.83$. 
     In all {\it CosmoRuns} we analyzed, the median value of $M_{\rm halo} / M_{\rm DMO}$ is less than 0.83. 
     This means that the halos originated from the same patch in the initial condition but under the influence of baryonic physics did not grow as much in mass as their DMO counterparts did.      
     See Section \ref{matching_pairs} for more information. 
     }
    \label{fig:matching_mass}
    \vspace{2mm}
\end{figure*}

\begin{itemize}

\item  As early as at $z = 12$, the {\it CosmoRuns} tend to have fewer halos than the DMO runs, even before the cosmic reionization begins or the extragalactic ultraviolet background radiation is turned on in {\sc Grackle}.\footnote{We set the {\sc Grackle} parameters {\tt UVbackground\_redshift\_on} $= 15$ and  {\tt UVbackground\_redshift\_fullon} $= 15$.}  
It is especially true in the higher-mass bin (right panels; $M_{\rm halo} > 10^8 h^{-1}  \msun$).
At $z \sim 15$, smaller dark matter halos have difficulties at keeping baryons because the density fluctuation of gas is smoother than that of dark matter on small scales \citep{1998MNRAS.296...44G}. 
Thus, these smaller halos have a smaller enclosed baryon mass than the cosmic average  \citep{2012ApJ...760....4O}. 
This leads to lower halo masses in the {\it CosmoRuns} at $z=15$, therefore, fewer halos in Figure \ref{fig:num_halo_by_z}.  
From $z=15$ to 8 the difference between the {\it CosmoRuns} and the DMO runs persists in both mass bins.

\item Again as early as at $z = 12$, one can observe a discrepancy in the {\it CosmoRuns} between particle-based codes and mesh-based codes in the lower-mass bin (top left panel; $ 10^7 h^{-1} \msun  < M_{\rm halo} < 10^8 h^{-1} \msun$).  
Particle-based codes show a tendency to have more halos than mesh-based codes do until $z\sim6$ when {\sc Ramses} begins to behave like particle-based codes.
It is well documented that the particle-based codes may produce more satellite halos due to the so-called ``gas -- dark matter particle coupling'' in the early universe \citep{2003MNRAS.344..481Y, 2012ApJ...760....4O}. 
When a gas particle is close to a nearby dark matter particle, the gas particle could be captured in the  dark matter particle's potential well.   
This gas particle now obtains an artificial velocity that follows that of the dark matter particle, resulting in an increased power on small scales.  
Even a minute gas -- dark matter two-particle coupling could be a source of numerically-driven fluctuation, particularly in the early universe that is nearly homogeneous. 
While this artifact may be alleviated with {\it adaptive} gravitational softening, the particle-based codes in the {\it AGORA} {\it CosmoRun} suite adopted a fixed gravitational softening length (see Section \ref{sec:methodology}), prone to overproduction of satellite halos.\footnote{A new type of cosmological initial conditions generated with a higher-order Lagrangian perturbation theory may  provide another solution to this problem \citep{2021MNRAS.500..663M}.  It will enable us to start our simulation at $z \simeq 15$, much later than $z=100$ as in the {\it CosmoRun}, bypassing the gas -- dark matter coupling problem at high redshift entirely.} 
On the other hand, some DMO runs in mesh-based codes may have coarser force resolution at high redshift as they attempt to resolve small density fluctuations in the outskirts of the target halo \citep{2005ApJS..160....1O, 2008CS&D....1a5003H}, leading to smaller numbers of halos in e.g., {\sc ARt-I-dmo} and {\sc Ramses-dmo}.  

\item From $z=8$ to 4, reionization plays an important role in suppressing the growth of satellite halos in the {\it CosmoRun} (see Section \ref{sec:methodology} and Paper III), when other local baryonic physics mechanisms are yet to become effective. 
The extragalactic photoionizing background radiation heats and removes the gas prior to infall, and efficiently inhibits the growth of halos at  $z\lesssim8$  \citep{2015MNRAS.448.2941S, 2017MNRAS.467.1678Q}.\footnote{It is worth to remind the readers that, in DMO runs, the mass of the gas is included in the dark matter component, effectively making $\Omega_{\rm matter} = 0.272 = \Omega_{\rm DM}$ (see Section \ref{dmorun}).  Therefore, while the gas experiences hydrodynamic forces such as reionizing radiation and may ``evaporate'' in the {\it CosmoRun}, the gas mass contribute in whole to the growth of the halo in the DMO run.} 
Reionization is relatively more effective on the low-mass halos \citep[$v_{\rm circ,\,max} < 20 \,{\rm km\,s^{-1}}$;][]{2015MNRAS.448.2941S, 2016MNRAS.458.1559Z}.

\item At later times, other baryonic effects enhance the depletion of substructures when compared to the DMO counterparts.  
Gas in low-mass halos is removed by ram-pressure stripping before the infall, along with the extragalactic radiation field.  
Tidal stripping in the steep gravitational potential of the host halo becomes important now, and significantly affects the satellite halo population, especially in the intermediate-mass range \citep[$20 \,{\rm km\,s^{-1}}< v_{\rm circ,\, max} < 35 \,{\rm km\,s^{-1}}$;][]{2010ApJ...709.1138D, 2013ApJ...765...22B, 2014ApJ...786...87B, 2016MNRAS.456...85S, 2016MNRAS.458.1559Z, 2017MNRAS.467.4383S, 2017MNRAS.471.1709G,2019MNRAS.487.4409K}. However, tidal disruption induced by stellar bulge and disk in cosmological simulations could be overestimated due to insufficient resolution \citep[see][]{2020MNRAS.499..116W, 2022MNRAS.509.2624G}.

Stellar feedback such as supernovae also expels the gas and impedes the halos' mass growths  \citep{2013ApJ...765...22B, 2013ApJ...766...56M, 2014MNRAS.442.2641V, 2015MNRAS.451.1247S, 2017MNRAS.471.3547F}. 
These late-time baryonic processes can explain the widening gap between the {\it CosmoRuns} and the DMO runs at $z\lesssim4$ in both left and right panels of Figure \ref{fig:num_halo_by_z}.

\end{itemize}

In summary, we find that baryonic processes cause all hydrodynamic {\it CosmoRun} simulations to have fewer satellite halos than the DMO runs at nearly all redshifts. 
While baryonic physics left only indirect signatures in the halos' growth histories of {\it dark matter} masses ($M_{\rm halo}$), in Section \ref{stellar_halos} we will see its more direct impact on the halos' {\it stellar} components.

\vspace{3mm}
\subsection{How Baryonic Physics Affects Each Individual Halo}
\label{matching_pairs}

Until now we have mainly focused on the population of satellite halos, and how it changes with the inclusion of baryonic physics.   
To investigate how each individual halo is actually affected by the baryonic processes, we now match and compare halos in hydrodynamic simulations (e.g., {\sc Enzo} {\it CosmoRun}) to their counterparts in the DMO simulation (e.g., {\sc Enzo-dmo} run). 
The matching process is adapted from \cite{2015MNRAS.451.1247S} and \cite{2021MNRAS.tmp.2968L}. 
For every satellite halo in e.g., {\sc Enzo} {\it CosmoRun}, we first identify the 40 dark matter particles that are closest to the halo's center.  
Since the particle IDs are shared by the {\sc Enzo} and {\sc Enzo-dmo} run, we can locate these 40 particles in the {\sc Enzo-dmo} run.   
Then we search for a halo containing 50\% or more of these counterpart particles.  
Finally, by carrying out the same procedure in reverse, another link is obtained --- i.e., first find the 40 most bound particles in the {\sc Enzo-dmo} run, and then locate these particles in the {\sc Enzo} {\it CosmoRun}.  
A pair of two halos that are {\it bijectively mapped} (bidirectionally connected) between the two simulations are considered as a ``matched'' pair.
Particle IDs are identically assigned in the initial conditions of {\sc Changa}, {\sc Gagdet-3}, {\sc Gear}, {\sc Arepo-t}, {\sc Gizmo} and their DMO counterparts, so we can similarly find matched pairs in between the two codes.
But because particle IDs in {\sc Art-I}, {\sc Ramses} and their DMO counterpart simulations are {\it not} identically assigned in their initial conditions, halos in these two simulations need to be matched with a different method based on the distribution of dark matter particles at $z=100$ (see Appendix \ref{appendix_ramses} for details).

We conjecture that various baryonic processes have slowed down the growth of halos in  hydrodynamic simulations compared to their DMO counterparts.  
To test this hypothesis, in Figure \ref{fig:matching_mass}, we plot the ratio of the dark matter mass of an individual halo in the {\it CosmoRun} to that of its matched DMO counterpart ($M_{\rm halo} / M_{\rm DMO}$).  
A few observations to note:

\begin{itemize}

\item In all eight {\it CosmoRuns} matched to their respective DMO counterpart, the ratio $M_{\rm halo} / M_{\rm DMO}$ is on average less than $\Omega_{\rm DM}/\Omega_{\rm matter} = 0.83$ (marked with solid horizontal lines in Figure \ref{fig:matching_mass}), where $\Omega_{\rm matter} = 0.272$ and $\Omega_{\rm DM}=0.227$ (see Section \ref{cosmorun}). 
If the halo in the {\it CosmoRun} had followed the identical mass growth history as that of its DMO counterpart, the dark matter mass of the {\it CosmoRun} halo, $M_{\rm halo}$, should have been $M_{\rm DMO} \times \Omega_{\rm DM}/\Omega_{\rm matter} = 0.83 \,M_{\rm DMO}$. 
The fact that the ratio lies below 0.83 means that the halos have smaller masses and smaller virial radii in all hydrodynamic simulations when compared with their DMO counterparts.  
Although the halos originated from the same patch in the initial condition, the ones under the influence of baryonic physics did not grow as much in mass as their DMO counterparts did.   

\item The baryonic effects are present at all redshifts, slightly more so at later times (i.e., the $M_{\rm halo} / M_{\rm DMO}$ ratio is smaller at $z \sim 2$ than at $z=12$). 
The baryonic effects are the combination of early- and late-time processes, such as reionization inhibiting the growth of small satellite halos with shallow gravitational wells, and the host halo's tidal field stripping the halos of existing gas, as discussed in Section \ref{over_time}.

\item Among the {\it CosmoRuns}, the mesh-based codes tend to have lower $M_{\rm halo} / M_{\rm DMO}$ values than the particle-based codes do in general, especially at low-mass end. 
The discrepancy between the two code groups is considerable already at $z=12$, which is in line with the over-abundance of satellite halos in particle-based codes discussed in Section \ref{over_time}. 
Furthermore, consistent with the patterns observed in Figures \ref{fig:mass_function_z2} and \ref{fig:num_halo_by_z}, {\sc Art-I} and {\sc Enzo} have smaller ratios compared to the other codes.

\end{itemize}

\begin{figure}
    \centering
    \vspace{1mm}    
    \includegraphics[width=0.78\linewidth]{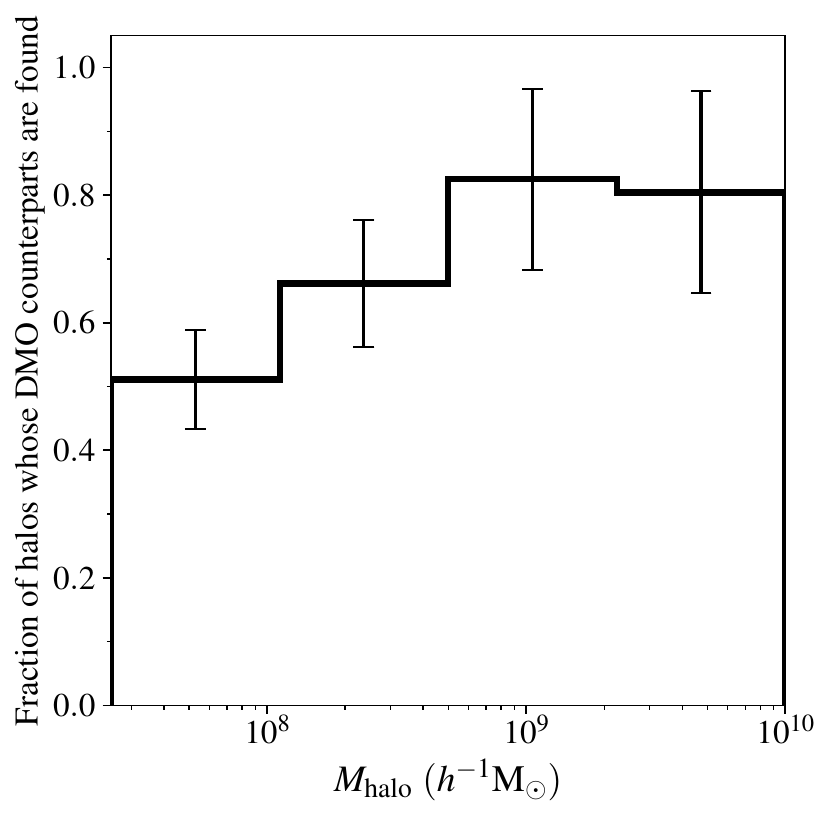}
    \vspace{-2mm}        
    \caption{The fraction of halos whose counterparts in the DMO simulations are identified at $z \sim 2$, averaged over all eight {\it CosmoRuns}.  
    Error bars indicate one standard deviation.
    The ``match fraction'' is lower for low-mass halos because the matching criteria is more demanding for them.  
    It implies that the low-mass matched halos are likely biased towards the ones with quiescent accretion histories.
    See Section \ref{matching_pairs} for more information. 
    }
    \label{fig:matched_fraction} 
    \vspace{0mm}    
\end{figure}

It should be noted that only a small fraction of halos are matched with their DMO counterparts in the low-mass end ($M_{\rm DMO} \lesssim 10^{8} h^{-1}  \msun$).
Figure \ref{fig:matched_fraction} illustrates the fraction of halos whose counterparts in the DMO simulations are identified at $z \sim 2$, averaged over all eight {\it CosmoRuns}. 
Since our halo matching process is more demanding for halos with fewer member particles, the ``match fraction'' is lower for low-mass halos. 
This suggests that the low-mass matched halos are likely biased towards the ones with quiescent accretion histories and without major mergers or disruptions in the past, potentially resulting in an overestimated $M_{\rm halo} / M_{\rm DMO}$ ratio.  
Readers may also find it interesting that massive satellite halos ($M_{\rm DMO} \gtrsim 10^9 h^{-1}  \msun$) have disappeared between $z=7$ and 4 (second and third panel from the right in Figure \ref{fig:matching_mass}). 
According to the accretion history of the host halo, multiple mergers occur between $z=7$ and 4, which explains the disappearance of the massive satellites by $z=4$.  

To summarize, we have shown that each individual halo tends to have a slower mass accretion history until $z\sim2$ in the {\it CosmoRun} than in its counterpart DMO run.
The discrepancy can be explained by early- and late-time baryonic physics that slows down the growth of satellite halos.

\begin{figure*}
    \vspace{1mm}
    \centering
    \includegraphics[width=0.88\linewidth]{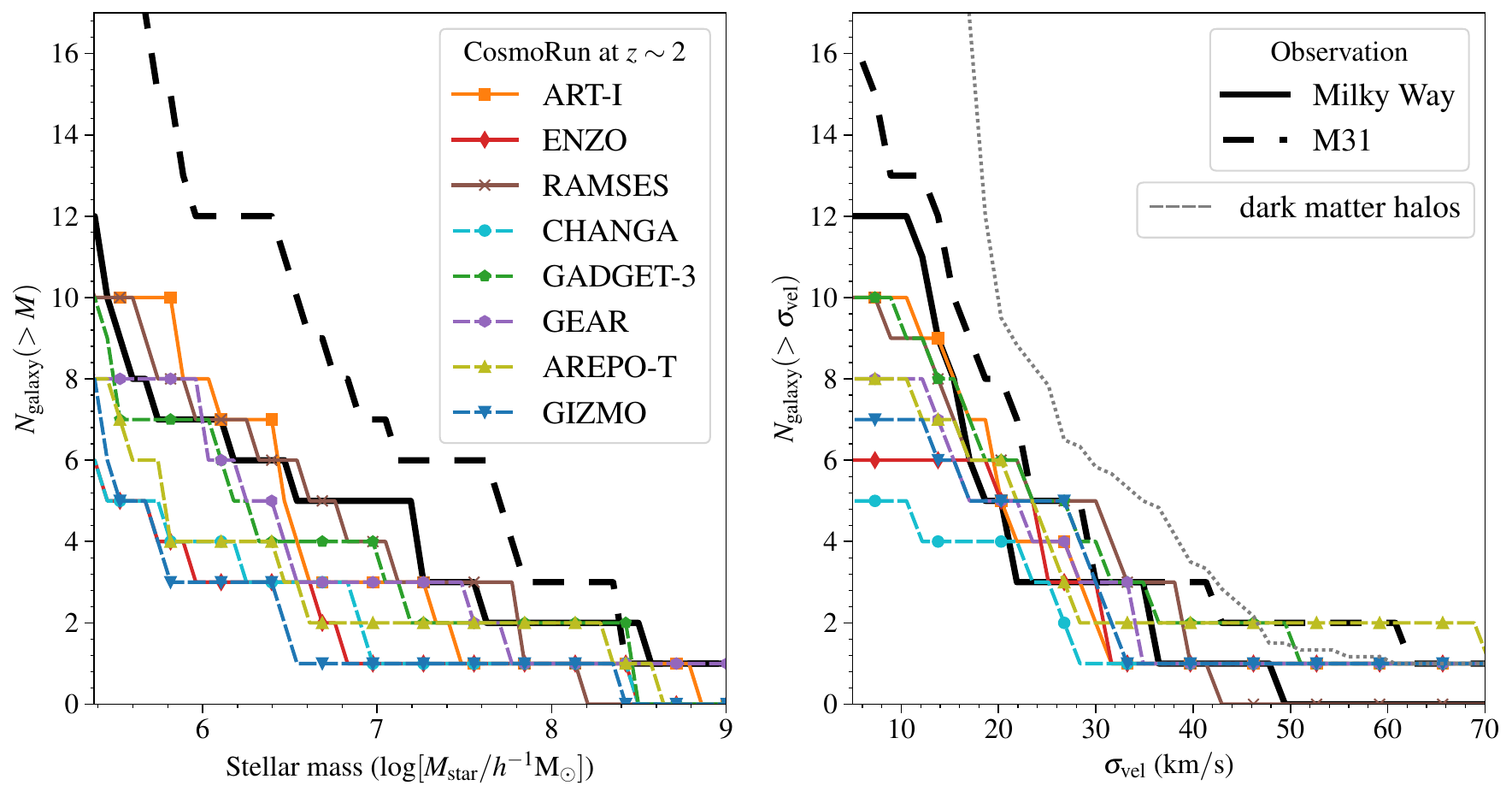}
    \vspace{-1mm}    
    \caption{The cumulative number of satellite {\it galaxies} at $z\sim 2$ in their stellar mass, $N_{{\rm galaxy}} (>M)$ ({\it left}), and in their 3-dimensional stellar velocity dispersion, $N_{{\rm galaxy}} (> \sigma_{\rm vel})$ ({\it right}). 
    We define galaxies as satellite halos that contain stellar particles.  
    The number of satellite {\it galaxies} is significantly fewer than that of satellite {\it halos} in all eight {\it CosmoRuns}. 
    Readers may compare $N_{\rm galaxy}$ with $N_{{\rm halo}}$ (in Figure \ref{fig:mass_function_z2}), or with the {\it gray dotted line} in the {\it right panel} above denoting the average number of satellite halos in all {\it CosmoRuns} (plotted with their dark matter velocity dispersion). 
    The {\it thick black solid line} and {\it dashed line} in both panels indicate the known present-day satellites around the Milky Way (MW) and M31, respectively, which of course are lower limits to the true numbers.
    See Section \ref{stellar_halos} for more information. 
    }
    \label{fig:starmass_function_z2}
    \vspace{3mm}    
\end{figure*}

\vspace{1mm}
\subsection{Satellite Galaxy Populations At $z \sim 2$}
\label{stellar_halos}

In Section \ref{abundance} we have demonstrated that all hydrodynamic {\it CosmoRun} simulations produce fewer satellite halos around our host halo than the DMO runs do across all satellite masses and radii.  
To verify that this finding naturally leads to the baryonic solution to the ``missing satellite problem'' (see Section \ref{abundance}) that is {\it independent of the numerical platform utilized}, in this section, we examine the satellite {\it galaxy} populations around the target host halo.
Here we define ``galaxies'' as satellite halos that contain stellar particles (see Section \ref{halo_finding} for more information).  

\begin{figure*}
    \vspace{1mm}
    \centering
    \includegraphics[width=0.9\linewidth]{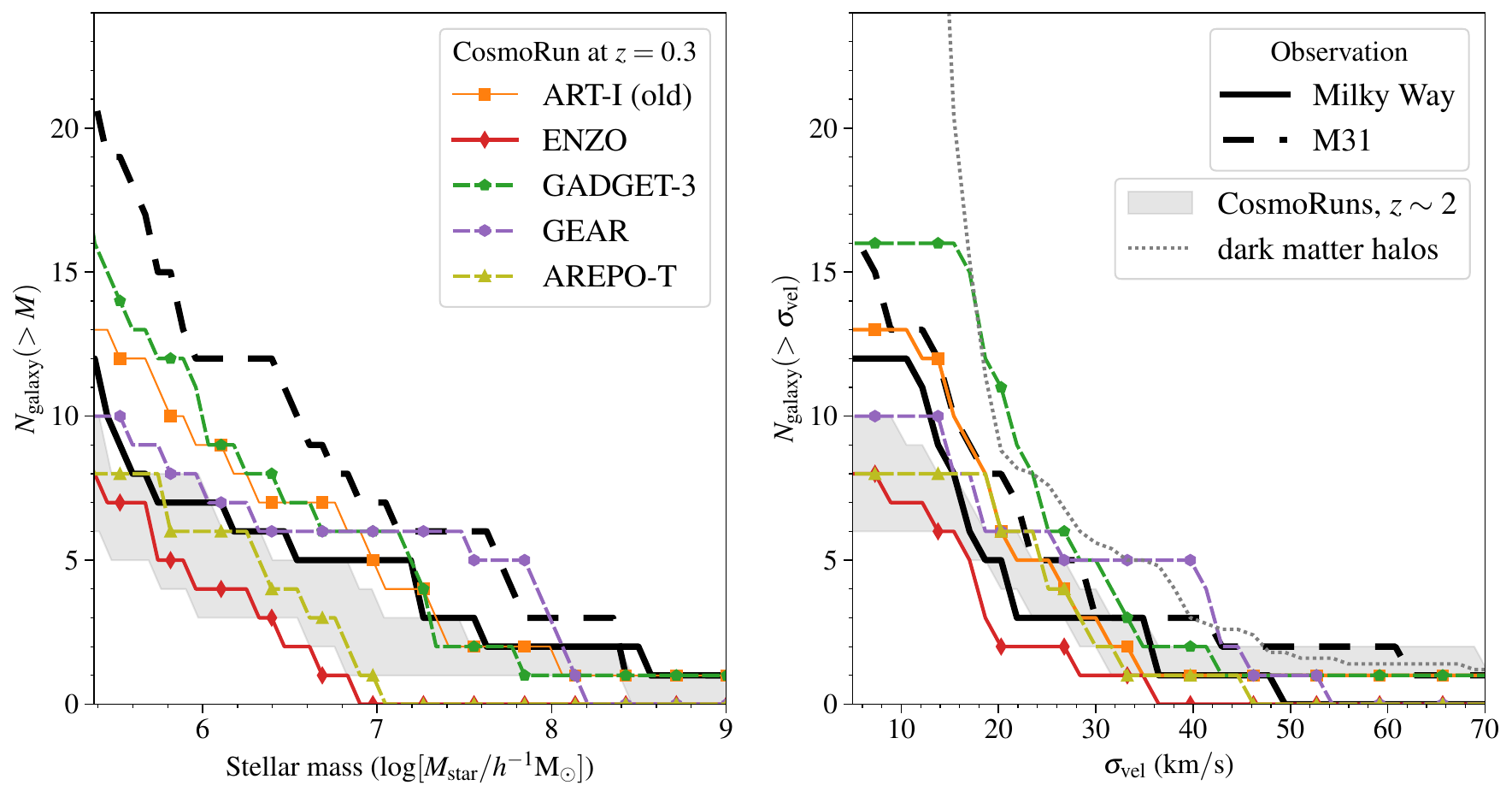}
    \vspace{-1mm}    
    \caption{ The cumulative number of satellite {\it galaxies} at $z=0.3$ in their stellar mass, $N_{{\rm galaxy}} (>M)$ ({\it left}), and in their 3-dimensional stellar velocity dispersion, $N_{{\rm galaxy}} (> \sigma_{\rm vel})$ ({\it right}), for the {\sc Art-I} (old), {\sc Enzo}, {\sc Gadget-3} and {\sc Gear} {\it CosmoRuns}. The grey area represents the minimum and maximum values of the cumulative number of satellite galaxies at $z\sim2$ from those codes. The numbers of satellite galaxies at $z=0.3$ are only a factor of $\lesssim 2$ larger than those at $z \sim 2$, or almost identical in abundance in some codes. As the latest {\sc Art-I} {\it CosmoRun} has not reached $z=0.3$, we use the older model for the {\sc Art-I} code, denoted as {\sc Art-I} (old), which is represented in Paper III. See Section \ref{galaxy_z0} for more information.}
    \label{fig:starmass_z0}
    \vspace{1mm}    
\end{figure*}

In Figure \ref{fig:starmass_function_z2} we plot the cumulative number of satellite galaxies at $z\sim 2$ in their stellar mass, $N_{{\rm galaxy}} (>M)$ (left), and in their 3-dimensional stellar velocity dispersion, $N_{{\rm galaxy}} (> \sigma_{\rm vel})$ (right). 
In both panels, we restrict the satellite galaxies to those with $M_{\rm star} > 6m_{\rm gas,\,IC} = 2.38\times 10^5 h^{-1}\msun$ (see Section \ref{halo_finding}). 
Several notable points are as follows: 

\begin{itemize}

\item By comparing with the mass function of satellite {\it halos}, $N_{{\rm halo}}$, in Figure \ref{fig:mass_function_z2}, or with the gray dotted line in the right panel of Figure \ref{fig:starmass_function_z2} denoting the average number of satellite {\it halos} in all {\it CosmoRuns}, one can readily see that the number of satellite {\it galaxies} is significantly fewer than that of satellite halos in all participating {\it CosmoRuns}.
While baryonic physics left indirect signatures in the halos' growth histories of {\it dark matter} masses in Sections  \ref{abundance} and \ref{over_time}, we can now see its more direct impact on the halos' {\it stellar} components.  
All the baryonic processes discussed in Section \ref{over_time} --- such as cosmic reionization, tidal stripping, ram pressure stripping, and stellar feedback --- act efficiently to halt or impede the stellar mass growth inside the halo \citep{2001ApJ...548...33B, 2018A&A...616A..96R}. 
For example, gas in a low-mass halo with a shallow potential well is removed by ram pressure stripping before its infall to the host halo, and by stellar feedback as supernovae explode.    
Further, these processes can interact; for example, supernova feedback can expel gas from halos which is then more easily removed by ram pressure stripping. 
As a result, gas is depleted in most low-mass satellite halos in the {\it CosmoRuns}, which end up with few stellar particles.

\item The thick black solid line and dashed line in both panels of Figure \ref{fig:starmass_function_z2} indicate the present-day satellites around the MW and M31, respectively \citep{2012AJ....144....4M}.\footnote{The latest compilation in 2021 can be found in \url{https://www.cadc-ccda.hia-iha.nrc-cnrc.gc.ca/en/community/nearby/}. For the stellar velocity dispersion of M33, however, we followed \cite{2022AJ....163..166Q}. To estimate the stelar mass of galaxies, we assume a mass-to-light ratio of 1.  Recent observations find more satellite galaxies in the Local Group, yielding 50$-$60 satellites around the MW.  However, most of these newly discovered galaxies are UFDs, which are fainter than $M_{\rm V} = - 7.7$ \citep[or $L = 10^5 \,{\rm L}_\odot$;][]{2019ARA&A..57..375S}.  Therefore, these newly discovered UFDs are beyond the scope of the present study owing to the limited numerical resolution.}$^{,}$\footnote{The 3-dimensional stellar velocity dispersions, $\sigma_{\rm vel}$, of the MW and M31 satellite galaxies are estimated by the line-of-sight stellar velocity dispersions multiplied by $\sqrt{3}$.  The number of satellite galaxies around M31 shown in two panels of Figure \ref{fig:starmass_function_z2} differ slightly due to the lack of stellar velocity information for two satellites.  That is, among the 19 observed satellites around M31, 16 have available stellar velocity information.}
Although readers should be cautioned that we are comparing two datasets at different epochs, one can observe that the satellite galaxy populations in the {\it CosmoRuns} at $z \sim 2$ are largely consistent with those of the MW and M31 at $z=0$ in their stellar masses and velocity dispersions. 
For more on how we attempt to compare the satellite galaxy populations at $z\sim0$, see Section \ref{galaxy_z0} and Figure \ref{fig:starmass_z0}.

\item The agreement amongst the satellite galaxy populations of the eight {\it CosmoRuns} is better in the right panel ($N_{{\rm galaxy}} (> \sigma_{\rm vel})$ in stellar velocity dispersion) than in the left panel ($N_{{\rm galaxy}} (>M)$ in stellar mass). 
It is because the velocity dispersion serves as a better and more useful proxy for the dynamical mass of a system, as it reflects the gravitational impact of the underlying dark matter halo (not just the stellar component of a galaxy). 
While the difference in the satellite galaxy population amongst the {\it CosmoRuns} is more pronounced when considering their stellar mass, there remains a good overall agreement. 
In Section \ref{stellar-to-halo}, we further investigate the relationship between the dark matter mass and stellar mass of the satellite halos.

\end{itemize}

To sum up, we have found that the number of satellite galaxies is significantly fewer than that of dark matter halos in all {\it CosmoRun} simulations, and is comparable to the number of present-day satellites near the MW or M31.  
The so-called ``missing satellite problem'' is resolved  {\it in all participating codes} simply by implementing the baryonic physics adopted for {\it AGORA} in simulations {\it with sufficient numerical resolution} ($\lesssim 100$ proper pc at $z=2$).
We argue that various baryonic processes make the {\it CosmoRuns} have far fewer satellite galaxies than the satellite dark matter halos in the DMO runs. 
Future studies tracing the star formation history and the trajectory of each halo will tell us which baryonic mechanism acts most prominently and when.

\vspace{3mm}
\section{Discussion}
\label{sec:discussion}

\subsection{Predicting Satellite Galaxy Populations At Lower Redshifts}
\label{galaxy_z0}

In Section \ref{sec:results}, we choose to study the satellite halo populations at $z\sim2$ because not all the {\it AGORA} {\it CosmoRuns} have yet reached $z=0$. This approach, of course, has limitations, as the majority of satellite halos at $z\sim2$ are likely to undergo mergers or be disrupted by $z=0$. Here we describe what form of satellite galaxy populations we expect to see at lower redshifts.  

In the DMO simulations, the number of satellite halos tends to increase over time, which may lead one to conclude that a higher satellite halo abundance is expected at $z\sim0$ than at $z\sim2$. 
However, as illustrated in Section \ref{over_time}, baryonic physics may disrupt halos, considerably reducing the number of satellite halos in the hydrodynamic simulations at lower redshift. 
And particularly because the {\it AGORA} {\it CosmoRun} adopted an initial condition of a halo with a quiescent merger history after $z=2$, the number of newly accreted satellite halos may be small after $z=2$. 
Therefore, we expect that the satellite galaxy population in the {\it CosmoRuns} would not change dramatically from $z\sim2$ to $z\sim0$. 

In order to verify this, in Figure \ref{fig:starmass_z0} we plot the satellite halo population at $z=0.3$ as a function of stellar mass and stellar velocity dispersion for the five codes that reached the epoch already.\footnote{{\sc Art-I} (old) in Figure \ref{fig:starmass_z0} represents the older {\sc Art-I} model used in Paper III. The satellite galaxy population at $z\sim2$ is almost identical between the two models. However, the older model, {\sc Art-I} (old), exhibits a more severe inter-code timing discrepancy.} 
The numbers of satellite galaxies at $z=0.3$ for these codes are either only a factor of $\lesssim 2$ larger than those at $z \sim 2$, or almost identical (as in the case of {\sc Enzo} and {\sc Arepo-t}). However the inter-code differences have greatly increased. For instance, {\sc Enzo} and {\sc Arepo-t} have no satellite galaxies with $M_{\rm star} > 10^7 h^{-1}\msun$ at that epoch, while {\sc Gadget-3} and {\sc Gear} each have six satellite galaxies exceeding that stellar mass. Compared by 3-dimensional stellar velocity dispersion, {\sc Enzo} has no satellite galaxies with $\sigma_{\rm vel} > 40$ km/s, while {\sc Gear} has five satellite galaxies exceeding that velocity dispersion. Despite these differences, the results still align well with the observed satellite galaxy populations for the Milky Way (MW) and M31, except that {\sc Enzo} shows a slightly lower population in both stellar mass and velocity dispersion, and {\sc Arepo-t} exhibits a reduction in stellar mass. We conclude that our findings in Section \ref{sec:results} for $z\sim2$ will likely also hold at $z\sim0$.  

\begin{figure*}
    \vspace{1mm}
    \centering
    \includegraphics[width=0.81\linewidth]{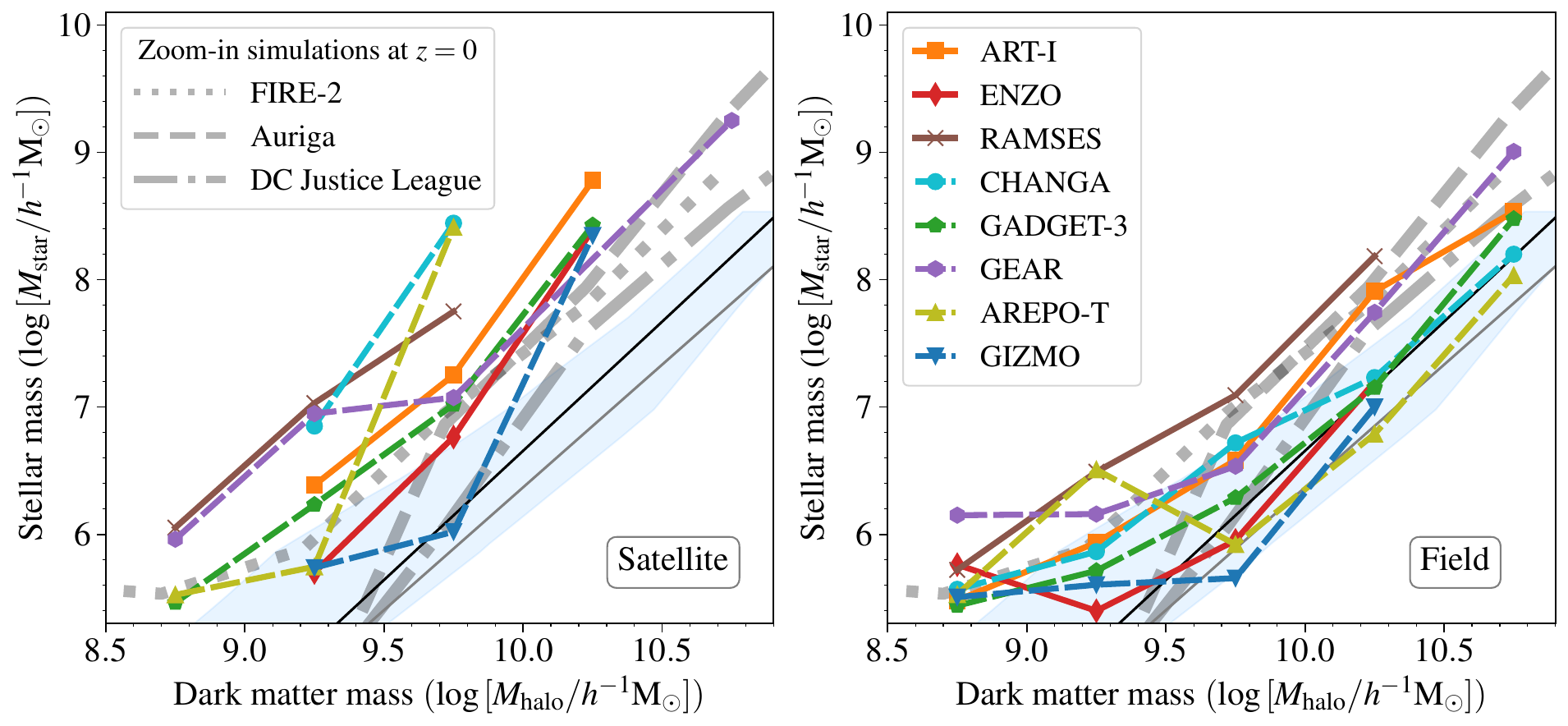}
    \vspace{0mm}    
    \caption{The stellar mass$-$halo mass relation of the satellite ({\it left}) and field ({\it right}) galaxies at $z\sim2$. 
    $y$-axis indicates the mean value of stellar masses in each mass bin.
    The {\it thick grey dotted}, {\it dashed} and {\it dot-dashed lines} are for dwarf galaxies in other zoom-in simulations at $z=0$ \citep[FIRE-2, Auriga and DC Justice League, respectively;][]{2018MNRAS.480..800H, 2021MNRAS.507.4953G, 2021ApJ...923...35M}, while the {\it thin black} and {\it grey solid lines} without markers are semi-empirical models for $2<z<2.5$ with extrapolation to low-mass galaxies \citep{2020A&A...634A.135G, 2019MNRAS.486.5468L}. The relationship with a 68\% confidence interval, as constrained by \cite{2020ApJ...893...48N} using Milky Way satellites, is represented in the blue shaded region. 
    All {\it CosmoRuns} produce similar relations with no systematic discrepancy between mesh-based codes and particle-based codes. 
    See Section \ref{stellar-to-halo} for more information.}
    \label{fig:smhm}
    \vspace{2mm}    
\end{figure*}

\begin{figure*}
    \vspace{1mm}
    \centering
    \includegraphics[width=0.81\linewidth]{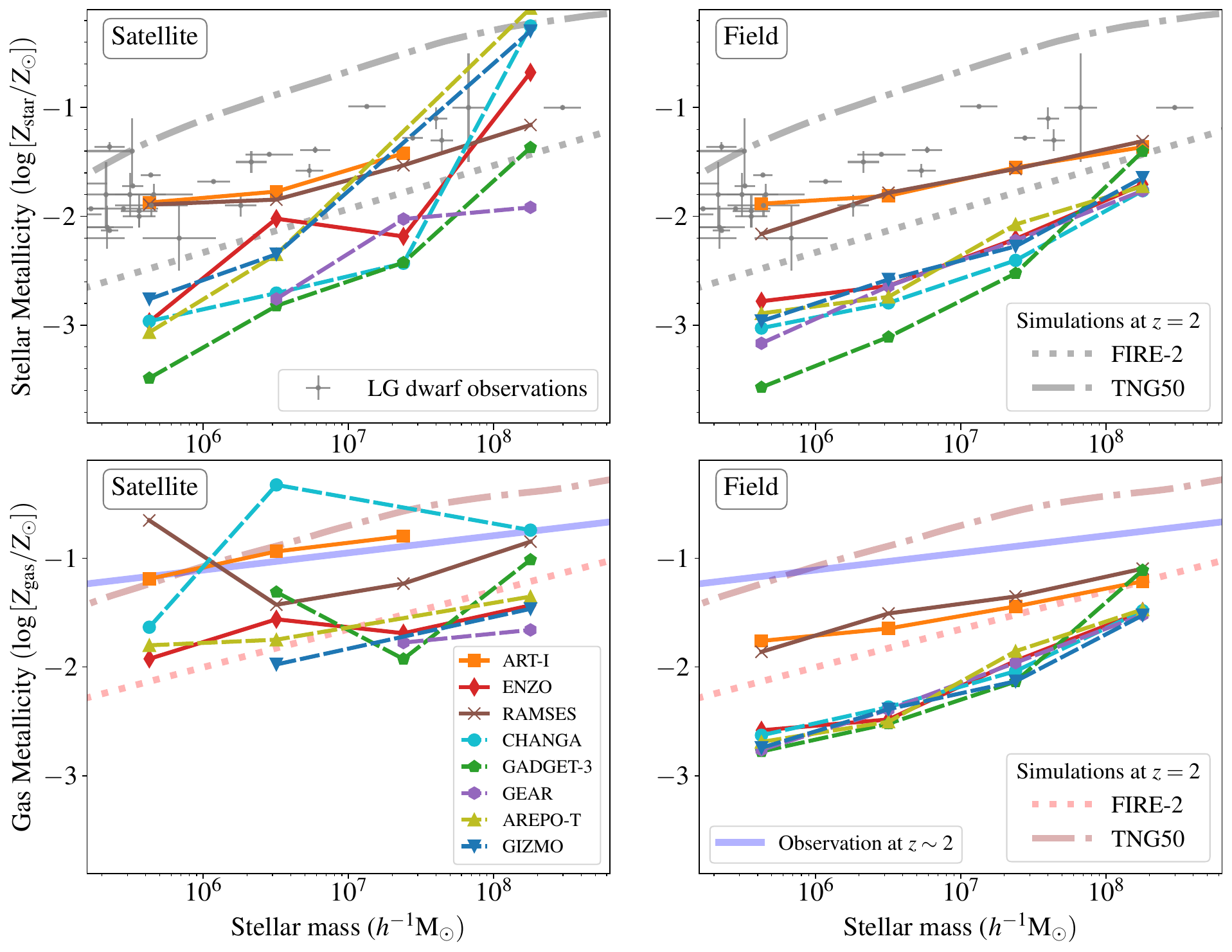}
    \vspace{0mm}    
    \caption{The mass$-$metallicity relation of the satellite ({\it left}) and field ({\it right}) galaxies at $z\sim 2$ using stellar ({\it top}) and gas ({\it bottom}) metallicity. 
    $y$-axis indicates the mean value of stellar metallicities in each mass bin. 
    A systematic difference exists between mesh-based codes ({\it solid lines}) and particle-based codes ({\it dashed lines}). 
    For comparison, the {\it thick grey (red) dotted line} represents the stellar mass$-$ stellar (gas) metallicity relation at $z=2$ that best fits the field galaxies in the FIRE-2 simulation \citep{2016MNRAS.456.2140M}, while the {\it thick grey (brown) dot-dashed line} represents the same relation at $z=2$ in the TNG50 simulation \citep{2019MNRAS.490.3196P, 2019MNRAS.490.3234N}. 
    We also include the observed mass-metallicity relation of dwarf galaxies in the Local Group, represented by {\it grey crosses} with error bars \citep[{\it top panels};][]{2023A&A...669A..94S}, as well as the mass-metallicity relation observed by JWST in a $z\sim2$ galaxy cluster field, shown as a {\it thick light blue line} \citep[{\it bottom panels};][]{2022arXiv221101382L}.
    See Section \ref{stellar-to-halo} for more information. 
    } 
    \label{fig:metal}
    \vspace{2mm}    
\end{figure*}

\vspace{1mm}
\subsection{Testing Inter-code Convergence In Satellite Properties:  The Stellar Mass$-$Halo Mass Relation And The Mass$-$Metallicity Relation} 
\label{stellar-to-halo}

We now explore the inter-platform convergence in satellite properties amongst the {\it CosmoRuns} by studying the two relations that probe the baryonic physics: the stellar mass$-$halo mass relation and the mass$-$metallicity relation.   
This will allows us to verify the realism of the {\it AGORA} baryonic physics in the {\it CosmoRun}.  

First, in the left panel of Figure \ref{fig:smhm}, we show the stellar mass$-$halo mass relation at $z\sim2$ of the satellite galaxies identified in Sections \ref{halo_finding} and \ref{stellar_halos}. For the completeness of our analysis, in the right panel of Figure \ref{fig:smhm},  we also draw the same plot using the {\it field galaxies} found in our simulations.
This is possible thanks to the sufficiently large zoom-in region around the host halo that contains 20$-$30 field galaxies at $z\sim2$.\footnote{In contrast to the satellite halos defined in Section \ref{halo_finding}, we define {\it field halos} using the following criteria: {\it (i)} a field halo must reside beyond 300 comoving kpc of our target host halo (or 100 proper kpc at $z=2$; a value similar to the virial radius of our host halo at $z=0$), {\it (ii)} it must be more massive than $10^7 h^{-1}\msun$ in dark matter, and {\it (iii)} it must not have a parent halo in the {\sc Rockstar} halo catalog (i.e., satellites of other halos are excluded). And after assigning stellar particles to these halos using the method described in Section \ref{halo_finding}, we plot only the {\it field galaxies} whose stellar masses are heavier than $6m_{\rm gas,\,IC} = 2.38\times 10^5 h^{-1}\msun$, just as in Section \ref{stellar_halos}.}  
One may notice that, on average, the dark matter halos of {\it field} galaxies are about 2.5 times more massive than the dark matter halos of {\it satellite} galaxies for a given luminosity.
The satellites' halo masses do not grow after their infall to the host, or rather, decrease due to tidal stripping.  
In the meantime, their stellar masses continue to grow \citep{1972ApJ...176....1G, 2019MNRAS.488.3143B}.
Thus, satellite galaxies tend to have more stellar masses at a given halo mass than field galaxies do.

Different simulation codes display varied behaviors, but there is also evidence of remarkable convergence. To begin, inter-code differences in the mass-metallicity relation are evident. {\sc Art-I}, {\sc Ramses}, and {\sc Gear} show a relatively large $M_{\rm star}/M_{\rm halo}$ value, while the ratios for {\sc Enzo} and {\sc Gizmo} are slightly smaller than those of other codes. This trend reflects what was already discovered in the satellite galaxy populations (left panel of Figure \ref{fig:starmass_function_z2}). {\sc Changa} also shows a large $M_{\rm star}/M_{\rm halo}$ for the satellite galaxies, which could arise from the insufficient number of satellite galaxies. The field galaxies in {\sc Changa} exhibit a relation consistent with other codes (right panel of Figure \ref{fig:smhm}).

However, despite these initial differences, the overall picture reveals convergence. The differences in stellar mass$-$halo mass relations are within 1 dex for the field galaxies across all {\it CosmoRuns} with no visible systematic discrepancy between mesh-based and particle-based codes. The common baryonic physics adopted in {\it AGORA} and the stellar feedback prescription typically used in each code group (calibrated to produce a similar stellar mass at $z=4$) are responsible for this convergence, particularly because the simulations are performed with sufficient resolution ($\lesssim 100$ proper pc at $z=2$). 

We then compare our result with previous studies. 
The thick grey dotted, dashed, and dot-dashed lines represent the relation for dwarf galaxies at $z=0$ in the FIRE-2, Auriga, and the DC Justice League simulation \citep[][respectively]{2018MNRAS.480..800H, 2021MNRAS.507.4953G, 2021ApJ...923...35M}.\footnote{These lines represent both satellite and field galaxies in both panels. In contrast to the previous studies listed here that employ the total halo mass, the halo mass $M_{\rm halo}$ in the present paper specifically refers to the mass of dark matter in the halo. However, as the majority of mass in satellite galaxies is dark matter, this slight difference in the mass definition does not significantly affect Figure \ref{fig:smhm}.}\footnote{\cite{2023MNRAS.519.3154H} show that the latest FIRE model, FIRE-3, predicts a stellar mass up to a factor of ten higher compared to the FIRE-2 model for dwarf galaxies with $M_{\rm peak} \approx 10^{9} \msun$. For the dwarf galaxies with  stellar masses $\gtrsim 10^6 - 10^7 \msun$, in contrast, there is little difference in galaxy stellar masses \citep{2023MNRAS.519.3154H}.} The blue shaded region represents the relation inferred from Milky Way satellites \citep{2020ApJ...893...48N}.
The thin black and grey solid lines are for the semi-empirical models at $2<z<2.5$ with extrapolation to dwarf-sized galaxies \citep[][respectively]{2020A&A...634A.135G, 2019MNRAS.486.5468L}. 
The stellar masses in the {\it AGORA} {\it CosmoRuns} are on average $\sim$ 0.5 dex higher than the empirical predictions, but are largely consistent with the previous simulation studies at $z=0$. 
The inter-code scatters in the low-mass halos ($M_{\rm halo} \lesssim 10^9 h^{-1} \msun$) is due to the complex interplay between baryonic physics and different merger history of the halos, which cannot easily be reproduced by abundance matching in the empirical models \citep{2018A&A...616A..96R}.\footnote{In Figure \ref{fig:smhm} some codes do not reach the highest dark matter mass bin, indicating an absence of satellite halos in that range. Similarly, some codes do not reach the lowest dark matter mass bin because there is no halo with a sufficient number of stellar particles in that mass range.}
The galaxy$-$halo connection seen in Figure \ref{fig:smhm} indicates not only the robustness and reproducibility of the participating simulations, but also the realism of the {\it AGORA} {\it CosmoRun} baryonic physics.

Second, in Figure \ref{fig:metal}  we present the mass$-$metallicity relation at $z\sim2$  for the satellite and field galaxies. 
Stellar masses and metallicities are used to draw the plots in the top panels, while stellar masses and {\it gas} metallicities are used in the bottom panels. 
For the bottom panels, we assign a gas parcel (cell or particle) to a halo if it is within $0.15 R_{\rm vir}$ from the halo's center. 
Only galaxies whose gas mass is greater than three times the approximate mass resolution of stellar particles, $3m_{\rm gas,\,IC}=1.19\times 10^5 h^{-1}\msun$ (see Section 3.1 of Paper III), are included in the bottom panels of Figure \ref{fig:metal}.

We note that there exists a small, but systematic difference between the mesh-based and particle-based codes. Some mesh-based codes, {\sc Art-I} and {\sc Ramses}, tend to show higher stellar metallicities than the particle-based codes do for the satellite galaxies (left panels in Figure \ref{fig:metal}), and a similar trend exists for the field galaxies, too (right panels). However, the differences between codes are mitigated the high stellar mass end, $M_{\rm star} \gtrsim 10^8\msun$, where the mean values of the relation converge within $\sim 0.5$ dex for the field galaxies (right panels).
Since our careful calibration of stellar feedback for the {\it CosmoRun} has yielded similar star formation histories across the participating codes (see Papers III and IV for detailed discussion), the difference in metallicities is most likely due to the difference in the metal transportation scheme each simulation has adopted. 
We have already reported a systematic discrepancy in the metal distribution between the two hydrodynamics approaches in the isolated galaxy comparison (Paper II).  
And \cite{2021ApJ...917...12S} quantified the difference in metal distribution caused by different metal diffusion schemes and different numerical resolutions (especially in galactic halos). 
We will further investigate the circumgalactic and intergalactic media of the {\it CosmoRuns}, providing clues to the origin of the discrepancy in their metal content.

Comparing the results with previous studies, the best fit to the FIRE-2 simulations sits right in the middle of our eight simulations \citep[thick  dotted lines;][]{2016MNRAS.456.2140M}, while that to the TNG50 simulation sits $\sim$ 1.0 dex higher in metallicity than our {\it CosmoRuns} do \citep[thick dot-dashed lines;][]{2019MNRAS.490.3196P, 2019MNRAS.490.3234N}. 
Additionally, we include the observed mass$-$metallicity relation of present-day dwarf galaxies in the Local Group \citep[grey crosses with error bars;][]{2023A&A...669A..94S}, and the median value of the relation for 29 galaxies in a galaxy cluster field at $z\sim2$ \citep[thick light blue line; JWST;][]{2022arXiv221101382L}.\footnote{We adopt $\log{(Z_{\rm gas}/Z_{\odot})} = 12 + \log{\rm (O/H)} - 9.0$ \citep{2016MNRAS.456.2140M}.} 
Both of these observations show $\sim 0.5$ dex higher metallicity than the satellite galaxies in the {\it CosmoRuns} do. 
Considering that the metallicity of dwarf galaxies tends to increase from $z=2$ to 0 (by $\sim0.4$ dex in the FIRE-2 simulations), this difference between the {\it CosmoRuns} and Local Group dwarfs may be less pronounced at $z=0$.
The mass$-$metallicity relation seen in Figure \ref{fig:metal} is an important test of the realism of the feedback prescriptions used in the {\it CosmoRuns}. While all {\it CosmoRuns} at $z\sim2$ reproduce the stellar masses of satellite galaxies similar to that of the MW and M31, the differences in their metallicities indicate that the baryon physics models implemented in the {\it CosmoRuns} have limitations.

\vspace{1mm}
\section{Conclusion}
\label{sec:conclusion}

We have studied the satellite halo populations near $z=2$ in the high-resolution cosmological zoom-in simulations carried out on eight widely-used astrophysical simulation codes ({\sc Art-I}, {\sc Enzo}, {\sc Ramses}, {\sc Changa}, {\sc Gadget-3}, {\sc Gear}, {\sc Arepo-t}, and {\sc Gizmo}) for the {\it AGORA} High-resolution Galaxy Simulations Comparison Project. 
We use different redshift epochs near $z=2$ for each code (``$z\sim 2$'') at which the eight {\it CosmoRuns} are in the same evolutionary stage in the target halo's merger history, in order to alleviate the timing discrepancy. 
Our key results are as follows:

\begin{itemize}

\item All hydrodynamic {\it CosmoRuns} have fewer satellite halos than the DMO runs do at $z\sim 2$ across all halo masses. 
The numbers of satellite halos in all {\it CosmoRuns} are fewer than those in the DMO runs by a factor of $\sim\,2$ for $M_{\rm halo} < 10^{8.5} h^{-1} \msun$ (Section \ref{abundance}).  

\item The difference between {\it CosmoRuns} and DMO runs exists as early as at $z=12$.  
The discrepancies in the early universe can be explained by the ``gas -- dark matter particle coupling'' in the particle-based codes and/or by the coarse force resolution in the mesh-based codes in the outskirts of the target halo. 
Other late-time baryonic effects such as reionization, tidal stripping, ram pressure stripping, and stellar feedback enhance the depletion of substructures when compared to the DMO counterparts (Sections \ref{over_time} and \ref{matching_pairs}).  

\item When we consider only the halos containing stellar particles at $z \sim 2$, the number of satellite {\it galaxies} is significantly fewer than that of dark matter halos in all participating {\it AGORA} simulations.
The populations of satellite galaxies in all eight {\it CosmoRuns} are indeed comparable to that of present-day satellites near the MW or M31 in their stellar masses and in their 3-dimensional stellar velocity dispersions.  
This finding is in line with previous studies \citep[Section \ref{stellar_halos}; see also][]{2014ApJ...786...87B, 2016MNRAS.457.1931S, 2016ApJ...827L..23W, 2021ApJ...906...96A}.

\item Using the five {\it CosmoRuns} that reached $z=0.3$, we also show that the number of satellite galaxies at $z=0.3$ are expected to be only a factor of $\lesssim 2$ larger than that at $z \sim 2$.
Thus, our conclusion that the number of satellite {\it galaxies} is significantly fewer than that of satellite {\it halos} will likely also hold at $z\sim0$ (Section \ref{galaxy_z0}).

\item We also find small, but systematic differences in other galaxy properties such as the stellar mass$-$halo mass relation and the mass$-$metallicity relation. {\sc Art-I}, {\sc Ramses}, and {\sc Gear} show a relatively large $M_{\rm star}/M_{\rm halo}$ value, while the ratios for {\sc Enzo} and {\sc Gizmo} are slightly smaller than those of the other codes. Similarly, {\sc Art-I} and {\sc Ramses} exhibit a relatively large mass$-$metallicity relation (Section \ref{stellar-to-halo}). We observe the differences in the metallicities between {\it CosmoRuns} and the observations, which indicate that the baryon physics models implemented in the {\it CosmoRuns} have limitations.
\end{itemize}

Overall, it is notable that the so-called ``missing satellite problem'' is fully and easily resolved across all participating codes simply by implementing the common baryonic physics adopted in {\it AGORA} and the stellar feedback prescription commonly used in each code group, with sufficient numerical resolution ($\lesssim 100$ proper pc at $z=2$). 
We have demonstrated that the baryonic solution to the decade-old problem in the $\Lambda$CDM model is effective in all eight {\it AGORA} participating codes at $z \sim2$. 
Because the results of our numerical experiment are reproduced by one another through the {\it AGORA} framework, the solution is {\it independent of the numerical platform adopted} --- excluding the possibility that it is an artifact of any one particular numerical implementation.  
Note that the stellar feedback prescriptions in the {\it CosmoRun} suite were calibrated to produce similar stellar masses in the host halo by $z=4$ (see Section 5.4  in Paper III) which remains true to $z \sim 2$ (see Paper IV), but they were never specifically aimed or designed to suppress the satellite galaxy population.


\vspace{2mm}

We thank all of our colleagues participating in the {\it AGORA} Project for their collaborative spirit which has enabled the Collaboration to remain strong as a platform to foster and launch multiple science-oriented comparison efforts.  
We particularly thank Oscar Agertz, Kirk Barrow, Oliver Hahn, Desika Narayanan, Eun-jin Shin, Britton Smith, Ben Tufeld and Matthew Turk for their insightful comments during the work presented in this paper, and Yongseok Jo and Seungjae Lee for their helpful feedback on the earlier version of this manuscript.  
We are also grateful to Volker Springel for making the {\sc Gadget-2} code public, and for providing the original versions of {\sc Gadget-3} to be used in the {\it AGORA} Project. 
This research used resources of the National Energy Research Scientific Computing Center, a DOE Office of Science User Facility supported by the Office of Science of the U.S. Department of Energy under Contract No. DE-AC02-05CH11231. 
Ji-hoon Kim acknowledges support by Samsung Science and Technology Foundation under Project Number SSTF-BA1802-04. 
His work was also supported by the National Research Foundation of Korea (NRF) grant funded by the Korea government (MSIT) (No. 2022M3K3A1093827 and 2023R1A2C1003244). 
His work was also supported by the National Institute of Supercomputing and Network/Korea Institute of Science and Technology Information with supercomputing resources including technical support, grants KSC-2020-CRE-0219, KSC-2021-CRE-0442 and KSC-2022-CRE-0355.
The publicly available {\sc Enzo} and {\tt yt} codes used in this work are the products of collaborative efforts by many independent scientists from numerous institutions around the world.  
Their commitment to open science has helped make this work possible.   


\appendix

\section{Halo Matching Process Between \texorpdfstring{\sc A\MakeLowercase{rt}-I}{Art-I}/\texorpdfstring{\sc R\MakeLowercase{amses}}{Ramses} and \texorpdfstring{\sc A\MakeLowercase{rt}-I-\MakeLowercase{dmo}}{Art-I-dmo}/\texorpdfstring{\sc R\MakeLowercase{amses-dmo}}{Ramses-dmo}  }
\label{appendix_ramses}

To investigate how each individual halo is affected by the baryonic processes, in Section \ref{matching_pairs} we match and compare halos in hydrodynamic simulations (e.g., {\sc Enzo} {\it CosmoRun}) to their counterparts in the DMO simulation (e.g., {\sc Enzo-dmo} run). 
But because particle IDs in {\sc Ramses} and {\sc Ramses-dmo} (or {\sc Art-I} and {\sc Art-I-dmo}) are not identically assigned in their initial conditions, we need to employ a method different from what is described in Section \ref{matching_pairs} to match the halos between these two simulations. 
Here we introduce an alternative approach to find a pair of matched halos between two simulations that only share the initial condition, but not their particle IDs. 
The idea is that we find a pair of halos originating from the same dark matter patch at a nearly homogenous early universe. 
First, we choose 40 particles closest to a target halo's center in e.g., {\sc Ramses} {\it CosmoRun} at $z\sim2$.  
We trace each dark matter particle in a halo back in time, and find its position at $z=100$ (the initial condition). 
Now for each of the 40 particles in the {\sc Ramses} run, a particle in the  {\sc Ramses-dmo}  run is randomly assigned.
For each pair of particles we can compute the distances between them at $z=100$.
Among the 40 edges, we sum up the 20 smallest distances.
A group of particles with the {\it smallest distance sum} is chosen in the {\sc Ramses-dmo} initial condition, and we trace them forward in time to find a ``matched'' halo at $z\sim2$.  
Finally, by carrying out the same procedure in reverse, another link is obtained --- i.e., first find the 40 most bound particles in the {\sc Ramses-dmo} run, and then locate their counterpart particles in the {\sc Ramses} {\it CosmoRun}.  
A pair of two halos that are {\it bijectively mapped} (bidirectionally connected) in between the two simulations are considered as a ``matched'' pair.

\bibliography{main}{}
\bibliographystyle{aasjournal}



\end{document}